\documentclass[12pts,
aps,
floatfix,
letterpaper,
prx,
longbibliography,
singlecolumn,
reprint,
superscriptaddress]{revtex4-2}
\usepackage[up,bf,raggedright]{titlesec}
\usepackage[sectionbib]{bibunits}
\usepackage{newtxtext,newtxmath}

\usepackage[colorlinks=true, linkcolor=blue]{hyperref}% add hypertext capabilities
\usepackage{lipsum}
\usepackage{graphicx} 
\usepackage{amsfonts}
\usepackage{amsmath}
\usepackage{braket}
\usepackage{gensymb}
\usepackage{hyperref}
\usepackage{dsfont}
\usepackage[capitalize]{cleveref}
\usepackage[utf8]{inputenc}

\date{\today}

\renewcommand{\thesection}{\arabic{section}}

\begin{document}

\title[High-quality nanostructured diamond membranes for nanoscale quantum sensing]{High-quality nanostructured diamond membranes for nanoscale quantum sensing}

\author{Alexander C Pakpour-Tabrizi}
\altaffiliation{These authors contributed equally to this work.}
\affiliation{Princeton University, Department of Electrical and Computer Engineering, Princeton, NJ, USA}
\author{Artur Lozovoi}
\altaffiliation{These authors contributed equally to this work.}
\affiliation{Princeton University, Department of Electrical and Computer Engineering, Princeton, NJ, USA}
\author{Sean Karg}
\altaffiliation{These authors contributed equally to this work.}
\affiliation{Princeton University, Department of Electrical and Computer Engineering, Princeton, NJ, USA}
\author{Tecla Bottinelli Montandon}
\affiliation{University of Basel, Department of Physics, Basel, Switzerland}
\author{Melody Leung}
\affiliation{Princeton University, Department of Electrical and Computer Engineering, Princeton, NJ, USA}
\author{Kai-Hung Cheng}
\affiliation{Princeton University, Department of Electrical and Computer Engineering, Princeton, NJ, USA}
\author{Nathalie P. de Leon}\email{npdeleon@princeton.edu}
\affiliation{Princeton University, Department of Electrical and Computer Engineering, Princeton, NJ, USA}

%%%%%%%%%%%%%%%%%%%%%%%%%%%%%%%%%%%%%%%%%%%%%%%%%%%%%%%%%
\maketitle
%%%% Abstract
\textbf{Deploying nitrogen vacancy (NV) centers in diamond as nanoscale quantum sensors for condensed matter and materials physics requires placing the NV centers close to the sensing target. One solution is to fabricate diamond nanostructures and integrate them with materials and devices. However, diamond etching and ion milling can introduce subsurface damage and surface defects that degrade the charge stability and spin coherence of NV centers near the surface. Here we report a procedure for fabricating low-damage nanostructured diamond membranes, and we show that this fabrication scheme preserves the optical and spin properties of state-of-the-art shallow NV center quantum sensors, within nanometers of the diamond surface, while providing significant photonic enhancement. Furthermore, we demonstrate a pick-and-place transfer method, which enables integration with diverse sensing targets.}

%%%%%%%%%%%%%%%%%%%%%%%%%%%%%%%%%%%%%%%%%%%%%%%%%%%%%%%%%%%%%%%%%%%%%%%%%%%%%%%%%%%%%%%%%%%%%%%%%%%%%%%%

\section{Introduction}\label{sec1}

NV centers in diamond have spin coherence times at the millisecond scale, optical spin initialization, and optical spin readout at room temperature and in ambient conditions, making them excellent qubits for nanoscale quantum sensing. \cite{casola2018probing,aslam2023quantum, Rovny_Rev} One major area of activity is using NV centers as probes for condensed matter physics, to sense charge transport, \cite{kolkowitz2015probing, andersen2019electron, vool2021imaging, palm2024observation} magnetic structure, \cite{huxter2023imaging, tan2024revealing, tschudin2024imaging} and critical dynamics. \cite{ziffer2024quantum, liu2025quantum} Achieving nanoscale resolution and high magnetic field sensitivity requires placing the NV centers close to the system of interest, and thus many of these demonstrations rely on either growing or transferring the material of interest on the diamond surface, \cite{kolkowitz2015probing, andersen2019electron} or fabricating diamond cantilevers for scanning probe magnetometry. \cite{maletinsky2012robust,scheidegger2022scanning} Both approaches are hampered by poor NV center spin and charge properties near the diamond surface,\cite{bluvstein2019identifying, yuan2020charge} which can be further degraded by surface and subsurface damage introduced during microfabrication and device integration. \cite{kolkowitz2015probing,leibold2025influence} Recently, low damage, high purity processing of bulk diamond substrates has been shown to enable NV centers within nanometers of the surface with spin coherence times exceeding 100 $\mu$s and no degradation to charge state stability. \cite{Sangtawesin_SurfNoise} Integrating such state-of-the-art sensors into a nanostructured material for device integration and photonic enhancement would enable a wide range of sensing applications, particularly for condensed matter physics.

There have been several advances in recent years in diamond nanofabrication, including reactive ion etching to define membranes, \cite{corazza2025homogeneous}, pillars\cite{Wrachtrup_pillar,Maletinsky_AFMCryo,scheidegger2022scanning} and nanobeams, \cite{Loncar_beams_fab,Yacoby_tribeams_chempot,Du_tribeams_Fe} and smart-cut techniques to fabricate diamond-on-insulator devices. \cite{Membrane_2_High, guo_direct-bonded_2024, ding_high-q_2024} An alternative method for realizing free-standing nanostructures is a modified version of the single crystal reactive etching and metallization (SCREAM) process, which was first developed for silicon \cite{SCREAM} and recently adapted to single crystal, monolithic diamond. \cite{Khanaliloo_highQ,Barclay_waveguide} In this process, anisotropic, top-down etching is followed by the creation of a sidewall hardmask and subsequent quasi-isotropic etching to undercut the desired structures. In the quasi-isotropic etching step, setting the RF bias to zero induces a highly diffuse, isotropic, plasma that etches different crystallographic directions of diamond at different rates. \cite{Yacoby_isodeepdive} This strategy has been used to fabricate microdisk resonators \cite{Barclay_300000} and diamond nanobeams. \cite{Barclay_waveguide,TeonoFibreCoupled, hanson_waveguide, brevoord_large-range_2025, Englund_phc} 

Here we apply the quasi-isotropic etch as part of a process to fabricate a membrane consisting of a dense array of nanobeams, with sacrificial tethers to the host substrate. To achieve a highly uniform undercut across microns of dense nanostructures, we design the device to reduce variations in the quasi-isotropic etch rate. We perform the etch at lower temperature and inductively coupled plasma reactive ion etching (ICP-RIE) power than in reference \cite{Barclay_300000}, reducing the etch rate and enabling control over the undercut geometry as well as the beam and tether thicknesses. In particular, the sacrificial tether geometry can be tuned to achieve robust suspension while still allowing mechanical separation. This enables device integration through pick-and-place transfer via micro-manipulator probes. By simultaneously ensuring that the surfaces have low contamination, low subsurface damage, and high quality oxygen termination, we fabricate nanobeams that host NV centers within nanometers of the surface that have state-of-the-art nanoscale sensing performance. We show that the spin coherence time, charge state stability, and charge dynamics of shallow NV centers in these nanobeams are in line with "typical best" bulk substrate processing. Furthermore, the nanobeam form factor enables collection efficiency enhancement by up to a factor of seven when integrated with a sapphire target, in agreement with photonic simulations. 

\begin{figure*}[t]
    \centering
    \includegraphics[width=\textwidth]{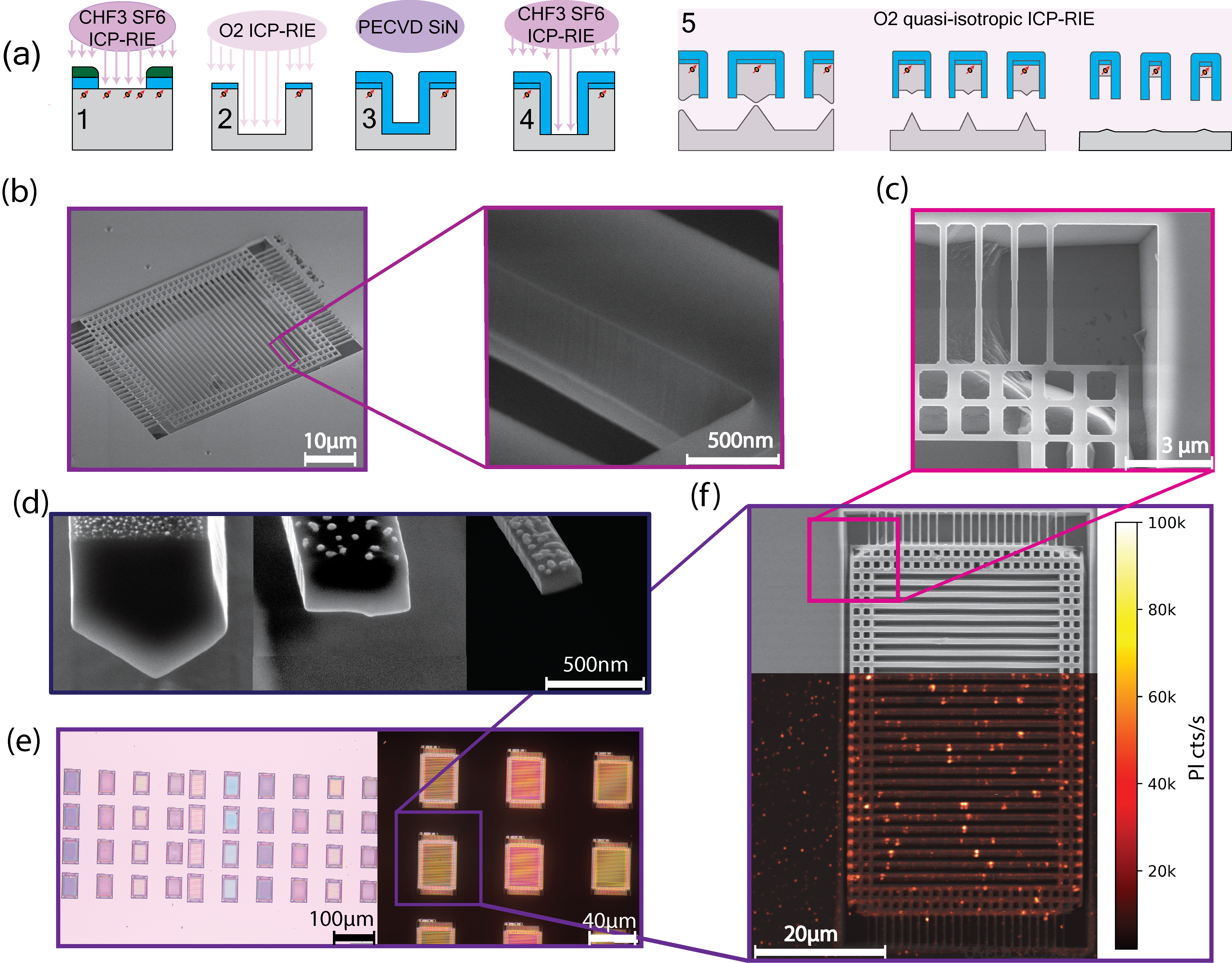}
    \caption{
        (a) Fabrication scheme showing method for fabricating freestanding undercut structures. Steps 1-4 comprise the anisotropic etch of the SiN$_x$ hardmask, the anisotropic etch of diamond, followed by conformal PECVD with SiN$_x$, then the final anisotropic etch of SiN$_x$ in the trench. Step 5 shows isotropic etching of three different width of nanobeams, resulting in different geometry of the underside. 
        (b) Tilted SEM image (left) of nanobeam membrane, with a higher magnification image (right) of an individual nanobeam (720 nm wide and 300 nm thick).
        (c) SEM image shows the design of the trellis, tethers and gaps between the frame and host sample. Thinning of the diamond on the outer trellis corner section can also been seen.
        (d) FIB cross-sections of three different gold-coated nanobeam widths fabricated in one device run. The high contrast beads on the top of the beams result from the gold coating.
        (e) Optical image of the host diamond sample showing many nanobeam membranes across a large area of the 4.5mm x 4.5mm host sample. 
        (f) SEM image overlaid with a confocal photoluminescence scan of a 720 nm wide nanobeam membrane acquired with air objective (NA = 0.9) under 520 nm (400 µW) illumination. Bright spots in the unfabricated host crystal and the nanobeams are NV centers.
    }
    \label{fig:Fab_fig1}
\end{figure*}

\section{Results}\label{sec2}

\subsection{Diamond Nanobeam Membrane Fabrication}

The fabrication process follows the approach of Mitchell et al. \cite{Barclay_300000}, with several adaptations to create high-density nanobeam membranes compatible with pick-and-place device integration for quantum sensing applications, as described below.
\\We prepare the surface of a (100) oriented electronic grade diamond (Element Six), according to the process outlined in \cite{Sangtawesin_SurfNoise} and implant it with $^{15}$N (3 keV, 1e9 cm$^{-2}$) followed by an 800$^\circ$C activation anneal in vacuum to create shallow NV centers. We then deposit 150~nm of silicon nitride (SiN$_x$) at 250$^\circ$C using plasma-enhanced chemical vapor deposition (PE-CVD, Oxford PlasmaPro 100), and pattern an etch mask (Zeon ZEP 520a) with electron beam lithography (Elionix ELS-F125), using a layer of spin-on conductive polymer (DisChem, DisCharge H2O X4) to reduce charging effects. The SiN$_x$ hard mask is patterned with a highly anisotropic etch (Figure~\ref{fig:Fab_fig1}a) using inductively-coupled plasma reactive ion etching (ICP RIE, PlasmaTherm Takachi SLR, 14 sccm CHF$_3$ and 10 sccm SF$_6$ at 5 mTorr, 200 W RF and 60 W ICP power). The high selectivity of the nitride in oxygen plasmas enables very smooth fabricated structures (Figure \ref{fig:Fab_fig1}b).
 
We first fabricate the membranes with variable nanobeam widths (120 - 720 nm) to study how nanobeam width and thickness influence the fabrication procedure. Based on finite-difference time-domain modeling (FDTD) (Figure \ref{fig:Supp1} and Section \ref{supp-fdtd}), we target $\approx$700 nm width and $\approx$350 nm thickness, which sets the dimensions for all the features in a given fabrication run. Due to the upward etch rate of the quasi-isotropic etch, we etch to a deeper depth than the desired thickness during the anisotropic diamond etch, balancing these two etch rates to achieve the final desired thickness. Etching in pure oxygen plasma (10 mTorr, 110 W ICP-RIE, 1000 W RF Bias) for 165 s removes 820 nm of diamond with near vertical side walls, no edge collapse, and minimal micromasking (Figure \ref{fig:rate_1}). Following the anisotropic etch, we deposit another 150 nm of SiN$_x$ conformally over the etched, SiN$_x$-capped diamond mesa. The residual SiN$_x$ hard mask is thinned during the anisotropic etch step but is still continuous. The hard mask on the bottom diamond surface between the mesas is selectively removed using another directional SiN$_x$ etch process, leaving a SiN$_x$ hard mask layer on the sidewalls and top surface. Crucially, the resulting encapsulation of the corners of the device is sufficient to protect the top edge of the membrane during the quasi-isotropic etch. 
 
The critical step is the long, quasi-isotropic etch performed in the same ICP-RIE with the platen heated to 140 °C. After an initial RF strike, the etch proceeds for 20 hours with 1000 W ICP and 0 W RF at 20 mTorr, resulting in quasi-isotropic etching along multiple directions. The lateral etch rate is strongly related to the distance between structures, as the quasi-isotropic etch rate is highly aspect-ratio dependent, necessitating long etches to achieve full undercut. For a 1 µm spacing, we find that the quasi-isotropic etch etches $\approx$600 nm laterally in 20 hours, guaranteeing full undercut for the targeted nanobeam width range. This aspect-ratio dependent rate is much smaller than in previous reports for structures such as microdisks, e.g. 2000 nm in 10 hrs ,\cite{Khanaliloo_highQ} and of sparse density rectangular photonic crystal cavities, \cite{Englund_phc} which do not suffer from an aspect ratio throttled etch rate. 
 
We use focused ion beam (FIB) milling to examine the cross section of nanobeams of different widths that are etched at the same time (Figure~\ref{fig:Fab_fig1}d). The surface texture arises from gold and iridium coating to mitigate charging during milling. The thinner nanobeams effectively etch for longer, resulting in their bottom profile progressing from faceted to flattened over etching time. The asymmetry in the nanobeam profiles is due to wafer edge effects, where the lateral etch rate from the quasi-isotropic etch can vary across the sample due to proximity of the nearest wafer edge.

Another key aspect enabling the pick-and-place operation is careful design of the outer frame (“trellis”) and attachment tethers. The double-layer trellis remains rigid, enabling transferable nanobeam frames with individual nanobeams as narrow as 100 nm and tens of micrometers long. The standardized trellis results in a high yield process, enabling consistent undercutting of thousands of membranes across a 4.5 x 4.5 mm$^2$ sample(Figure~\ref{fig:Fab_fig1}e). The 1 µm gap between the frame and host crystal ensures that the etch rate of the nanobeams is uniform even at the edges of the pattern, and the trellis buffers against aspect-ratio dependent edge effects, particularly at the  corners (Figure~\ref{fig:Fab_fig1}c) where there is significantly faster upwards etching during the quasi-isotropic step. The tethers are very narrow (\textasciitilde100 nm) and thin to ensure mechanical flexibility\cite{Suresh_mechFlex}, and we pattern a large number to build a mechanically reliable anchor. These tethers can be broken individually by micromanipulators while allowing the membranes to withstand harsh processing. Without this design, frames would collapse or delaminate, precluding scalable, high-yield fabrication. 

After isotropic etching, the sample is immersed in 49\% HF for 24 hours to strip the SiN$_x$, followed by cleaning in a refluxing mixture of nitric, perchloric, and sulfuric acids. Microscope images show that the resulting etched structures are smooth, and that the fabrication process has a high yield across a diamond wafer (Figure~\ref{fig:Fab_fig1}b,c,e). 

We use X-ray photoelectron spectroscopy to verify surface cleanliness prior to performing a thermal oxygen anneal at 450$^\circ$C according to the procedure described in \cite{Sangtawesin_SurfNoise}, which ensures high-quality oxygen termination. We image the resulting NV centers in a confocal microscope (NA = 0.9) under 520 nm illumination. The confocal microscope image overlaid with a scanning electron microscopy (SEM) image of the devices shows bright, single NV center emission in the nanobeams (Figure~\ref{fig:Fab_fig1}f). Single NV centers are brighter in the nanobeams compared to the host diamond, while their density is similar.

\begin{figure}
    \centering
    \includegraphics[width=1\linewidth]{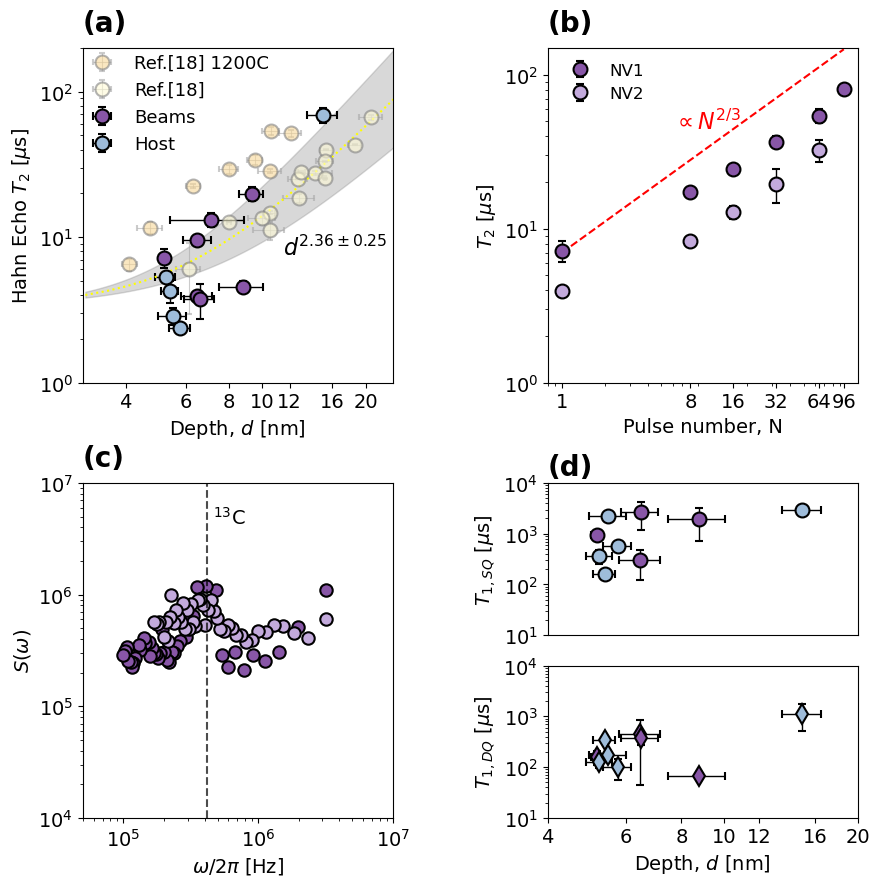}
    \caption{(a) Hahn echo coherence time T$_2$ as a function of NV center depth in nanobeams and the host diamond, plotted together with data from Ref \cite{Sangtawesin_SurfNoise} for shallow NV centers after the oxygen surface termination procedure including a 1200$^\circ$C and only a 800$^\circ$C anneal (as in the current work). Yellow dashed line shows a $T_2 = kd^n+T_{2,0}$ fit to the 800$^\circ$C annealed data from Ref. \cite{Sangtawesin_SurfNoise} ($k = 0.047\pm0.04, n = 2.35\pm0.25,  T_{2,0} = 3.35\pm3.87$). Grey shaded area marks the error margin for $n$ with fixed $k = 0.047$ and $T_{2,0}=3.35$ showing that most of the data that we measure in the nanobeams falls within this scaling range. (b) T$_2$ measured using an XY-N pulse sequence as a function of the number of pulses, N, (N = 1 is a Hahn echo experiment) for two NV centers in the nanobeams. Dashed line shows $\propto{N^{2/3}}$ dependence. (c) Spectral noise density extracted from a XY-8 decoherence decay using Equation S\ref{eqS2} for the same NV centers as in (b). The expected peak position associated with the  $^{13}$C Larmor frequency is marked with a vertical dashed line. (d) $T_{1,SQ}$ and $T_{1,DQ}$ as a function of NV center depth in nanobeams (purple) and unfabricated diamond (blue). }
    \label{fig:Figure2} 
\end{figure}

\subsection{NV Center Spin Properties}

We examine the spin properties of NV centers implanted within nanometers of the diamond surface in nanobeams to evaluate their utility as nanoscale quantum sensors, and we compare them to those in the unfabricated regions of the diamond. We measure the Hahn echo coherence time ($T_2$) as a function of depth from the surface for several NV centers in the nanobeams and in the unfabricated diamond (Figure \ref{fig:Figure2}a). The depth is measured using proton nuclear magnetic resonance, as has been described previously. \cite{pham2016nmr} The $T_2$ decreases as NV centers are brought closer to the diamond surface, as has been reported in the literature. \cite{Sangtawesin_SurfNoise,myers2014probing} The observed depth dependence of $T_2$ for NV centers in nanobeams is within the distribution of NV centers in the unfabricated regions, and is also consistent with state-of-the-art surface termination as described in the literature. \cite{Sangtawesin_SurfNoise} We note that the coherence times are slightly shorter than the best reported NV centers (Figure \ref{fig:Figure2}a, orange points), which resulted from a 1200$^\circ$C anneal that removes multivacancy clusters that form because of implantation damage, pointing to a potential direction for future improvement in the NV center performance in nanobeams.

To probe higher frequency noise, we measure the spin coherence with dynamical decoupling by XY-N pulse sequences. The coherence time $T_2$ increases with the number of pulses in the sequence, $N$, for two randomly selected NV centers (Figure \ref{fig:Figure2}b). We do not observe any saturation in $T_2$ up to 96 pulses, however the scaling deviates from the ${N^{2/3}}$ scaling expected for a slowly fluctuating noise bath,\cite{bar2012suppression} which can be attributed to the presence of broadband noise induced by defects at the diamond surface. Spectral noise decomposition using an XY-8 pulse sequence\cite{romach2015spectroscopy} for the same two NV centers reveals a noise spectrum $S(\omega)$ consistent with a broadband noise bath and a pronounced peak corresponding to the $^{13}$C nuclear spin Larmor precession at 360 G (Figure \ref{fig:Figure2}c).  These observations are consistent with previous characterization of the surface noise bath in state-of-the-art shallow NV centers. \cite{Sangtawesin_SurfNoise} 

Finally, we measure single-quantum spin lifetime ($T_{1,SQ}$) and double-quantum spin lifetime ($T_{1,DQ}$) to probe magnetic and electric noise near the NV center spin resonance frequency ($\approx$ 2.8 GHz)\cite{myers2014probing}, correspondingly, as a function of distance from the surface (Figure \ref{fig:Figure2}d). The spin lifetimes display a large distribution from 100 µs to several ms, with no clear depth dependence, and the distributions between NV centers in nanobeams and in the unfabricated region are similar. The $T_{1,DQ}$ values between 100 µs and 1 ms are similar to the results presented in \cite{Sangtawesin_SurfNoise}, which confirms that the fabrication process does not contribute additional sources of electric noise from the diamond surface. At the same time, we observe several shallow NV centers with $T_{1,SQ}$ in the range of 200 µs to 1 ms compared to $T_{1,SQ}$ > 1 ms as reported in \cite{Sangtawesin_SurfNoise}, which could point to slightly increased high-frequency magnetic noise at the surface.

\begin{figure}[b]
    \centering
    \includegraphics[width=1\linewidth]{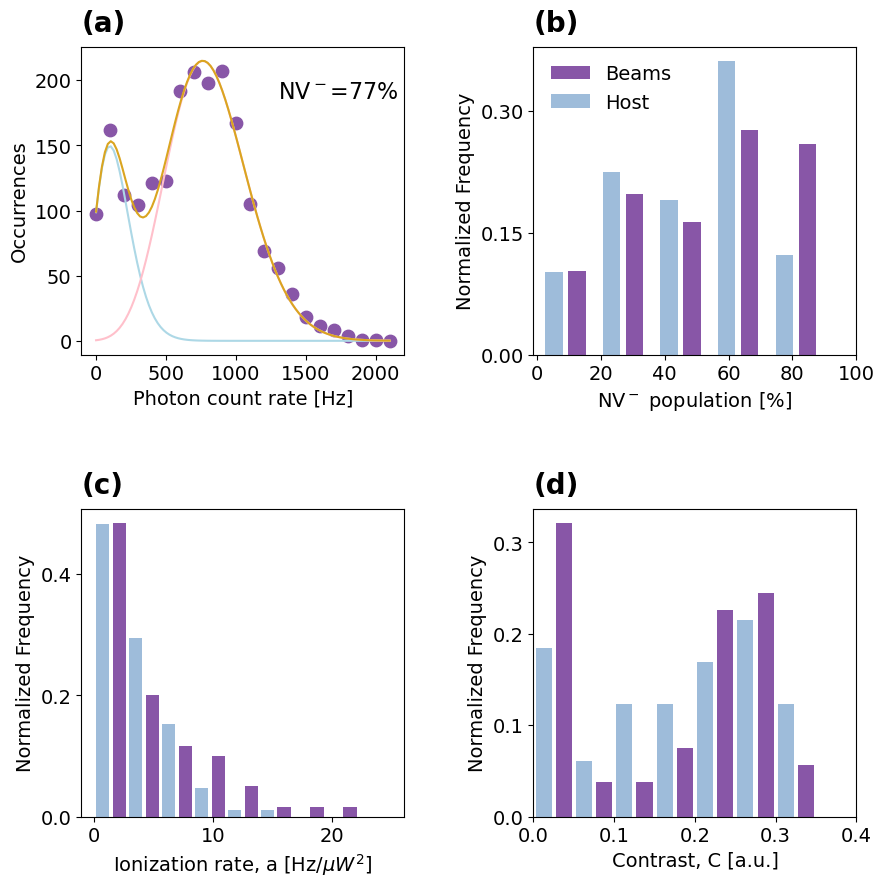}
    \caption{(a) Photon counting histogram of fluorescence from an NV center in a nanobeam under weak orange illumination ($\approx$3 $\mu$W) fitted with a double-Poisson distribution function reflecting the photon number distributions associated with the NV$^-$ (bright) or NV$^0$ (dark) charge states. This distribution corresponds to 77$\%$ NV$^-$ population. (b) Distribution of  NV$^-$ population measured based on the histograms as in (a) for NV centers in the nanobeams of various widths and the host diamond. (c) Distribution of  NV$^-$ ionization rate coefficients under orange illumination for NV centers in the nanobeams of various widths and for NV centers in the host diamond. (d) Distribution of  NV$^-$ spin contrast measured in a Rabi oscillation experiment at 25 G for NV centers in the nanobeams of various widths and for NV centers in the host diamond.} 
    \label{fig:Figure3}
\end{figure}

\subsection{NV Center Charge Properties}

Surface defects, impurities, and subsurface traps induced by etching may also degrade the NV center charge state properties, \cite{yuan2020charge, dhomkar2018charge} resulting in lower charge initialization or poor spin contrast. We measure the steady state charge population of NV centers in the nanobeams and in the host diamond by recording the photon number histogram under weak ($\approx$3 µW) 594 nm orange illumination after conventional green initialization (Figure \ref{fig:Figure3}a). A double Poissonian distribution fit reveals the two charge states: NV$^-$ (higher photon counts) and NV$^0$ (lower photon counts). We interrogate 116 NV centers in nanobeams of various widths and 148 NV centers in the host diamond and plot the resulting NV$^-$ population statistics in Figure \ref{fig:Figure3}b. The observed charge state distributions are very similar, with 50\% and 45\% of NV centers showing  >60\% NV$^-$ population in the nanobeams and the host diamond host, respectively. 

It has been reported that NV spin contrast can be compromised due to increased photoionization rates in the presence of imperfect surface termination even when the steady state NV$^-$ population is normal. \cite{yuan2020charge} Therefore, we measure the time-dependent photoluminescence under orange illumination of variable power and fit the individual photoluminescence decays to an exponential function to extract the corresponding ionization rates. The rates vary quadratically with power as $aP^2$, where $a$ is the ionization coefficient and $P$ is the power, as expected for two-photon photoionization and recombination (see Figure \ref{fig:Supp2s}). The statistics of the ionization coefficients $a$ for 61 NV centers in the nanobeams of various widths and for 85 NV centers in the host diamond are plotted in Figure \ref{fig:Figure3}c. We find the mean ionization rate coefficients to be 5.4 Hz/µW$^2$ and 3.6 Hz/µW$^2$ for nanobeams and for the host diamond, respectively, with $\approx$20\% of the NV centers in nanobeams displaying $a$ > 10 Hz/µW$^2$. This faster ionization can be attributed to the increased optical excitation efficiency in the nanobeams (discussed in more detail in Section \ref{subsec4}). 

To quantify the impact of this increased ionization rate on sensing performance, we measure the spin contrast by performing Rabi oscillations in a small magnetic field (25 G). The spin contrast is then calculated as $C = \frac{S_0 - S_1}{S_0}$ where $S_0$ ($S_1$) is the signal measured after initializing the ground state spin of the NV center into a $m_s = 0$ ($m_s = -1$) spin sublevel. The spin contrast distributions for 53 NV centers in the nanobeams of various widths and for 65 NV centers in the host diamond are shown in Figure \ref{fig:Figure3}d. 52\% and 56\% of NV centers in nanobeams and in the host diamond, respectively, are characterized by good spin contrast ($>$20\%). We observe more NV centers with no spin contrast in the nanobeams (around 50\% more), which we attribute to higher prevalence of NV centers with lower NV$^-$ population in this subset on the nanobeams (see Figure \ref{fig:Supp3s}). We conclude that the quality of the fabricated surfaces is compatible with adequate charge properties for nanoscale sensing.

\subsection{Device Integration and Photonic Optimization}\label{subsec4}

\begin{figure}
    \centering
    \includegraphics[width=1\linewidth]{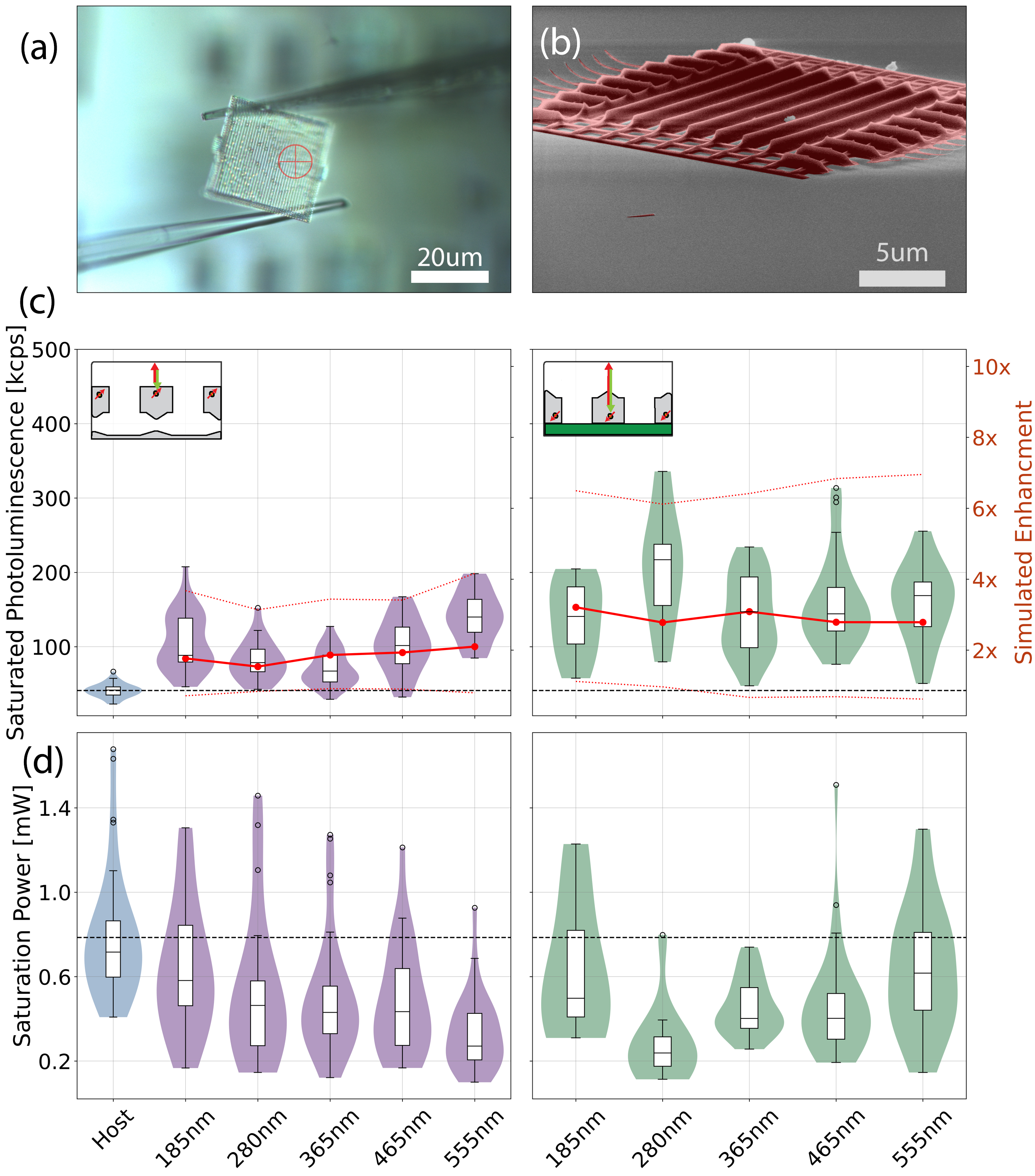}
    \caption{(a) Widefield image of a membrane mid-transfer, held electrostatically by two glass probes above the host diamond. (b) False-colored SEM of a 15um diamond membrane (red) fully transferred onto a sapphire substrate (grey). (c) Saturated photoluminescence intensity under 520 nm illumination for NV centers in the unfabricated host (left, blue) as compared to the nanobeams of various widths and thicknesses for suspended beams (left, purple) and beams transferred to a sapphire substrate (right, green). The solid red line represents the  enhancement in collection efficiency averaged over different NV dipole orientations and lateral displacement. Dashed red lines represent the maximum and minimum collection efficiency enhancement from simulations for a given nanobeam geometry. (d) Saturation power for NV centers in the unfabricated host (left, blue), suspended beams (left, purple), and beams transferred to a sapphire substrate (right, green). Dashed black line in (c) and (d) marks the average values in the unfabricated host, while the center box plot line represents the median of the data.}
    \label{fig:Figure4}
\end{figure}

It has been shown that pulled glass rods can be utilized for pick and place transfer of diamond cantilevers. \cite{Maletinsky_tran} To transfer our membranes, we use two glass micromanipulator probes, fabricated via micropipette pulling and mounted on motorized stages in a customized 2D-material transfer microscope (HQ Graphene). Using these probes, we first break the membrane tethers, then pick up the membranes, which are electrostatically held to the pipette tip. (Figure \ref{fig:Figure4}a). Using both manipulators, membranes are moved, flipped, and placed onto the surface of a sapphire substrate (Figure \ref{fig:Figure4}b). By applying mechanical pressure with a micromanipulator we can ensure that membranes make good contact with the substrate. For a clean sapphire surface, transferred membranes have sufficient contact to remain in place after wet processing with refluxing nitric, perchloric and sulfuric acid, as well as with HF, piranha solution (2:1 sulfuric acid and hydrogen peroxide), solvents such as acetone, IPA, and water, as well as blow drying under nitrogen. 

We measure the saturated NV center fluorescence rate and saturation power in nanobeams transferred onto sapphire and compare them to the NV centers in suspended nanobeams (Figure \ref{fig:Figure4}c). Both configurations experience an average collection efficiency enhancement over unfabricated host diamond, \textasciitilde2x in air and \textasciitilde3x on sapphire, with a large variation between the maximum and minimum possible values, up to a \textasciitilde7x improvement. We also observe a commensurate reduction in the saturation power for NV centers within nanobeams, corresponding to an increased excitation efficiency (Figure \ref{fig:Figure4}d), corroborating measurements of increased ionization rates in the nanobeams (Figure \ref{fig:Figure3}c).

FDTD simulations (Lumerical) are used to model the collection efficiency as a function of the nanobeam geometry, NV position, dipole orientation, and index of refraction of the environment. We approximate the nanobeam cross sections as perfect rectangles. Nitrogen implantation introduces stochastic variation in depth and lateral position of NV centers with respect to the nanobeam. We assume a 6 nm implantation depth without straggle, given a minor dependence of the collection efficiency on NV depth (Figure \ref{fig:Supp2}). There is a strong dependence of the collection efficiency on the lateral position of the NV center due to the proximity of the diamond sidewall-air interface. Furthermore, different dipole orientations, namely, the pair of dipoles oriented toward the objective, and the pair oriented away from the objective in (100) diamond, are brighter and darker, respectively. Consequently, the possible range of enhancement is large, with some points in this space experiencing no enhancement and some reaching up to a 7x improvement in collection efficiency. The maximum and minimum collection efficiency for a given nanobeam geometry is shown as a red-dashed line (Figure \ref{fig:Figure4}c), highlighting this spread. A detailed analysis of this variation can be found in the Supporting Information (Section \ref{supp-fdtd}). We perform this parameter sweep for two different sensing configurations: suspended nanobeams and nanobeams heterogeneously integrated (implanted-side down) onto a sapphire substrate. When flipped, the NV center position is close to the substrate, changing both the effective refractive index of the environment, as well as its relative position with respect to the objective and the nanobeam. This position benefits from total internal reflection at the nanobeam sidewalls, waveguiding the emission and enhancing the overall collection efficiency compared to the suspended case. We observe good agreement between the simulated and the observed enhancement in saturated fluorescence (Figure \ref{fig:Figure4}c). While we do not measure the precise lateral displacement beyond the optical diffraction limit, we observe variation in the collection efficiency that is consistent with the simulated distribution.

\section{Conclusions}\label{sec3}

We establish nanostructured diamond membranes that host state-of-the-art shallow NV centers as a practical platform for quantum sensing experiments. The membrane fabrication allows for long NV center spin coherence and stable charge dynamics, while structuring the membrane into nanobeams enables enhanced photon collection. We have demonstrated a device geometry that is compatible with polymer-free, pick-and-place mechanical transfer, which can be readily applied to sensing in a variety of superconducting, magnetic, and other condensed matter systems. In particular, this sensing geometry aids in the interrogation of optically opaque materials, which is a particularly challenging task for single-crystal diamond sensors. Additionally, enhanced collection efficiency and reduction in required excitation power improves measurement throughput and reduces optical illumination of the target material. Future potential extensions of these capabilities include patterned implantation of NV centers using a nanopatterned mask \cite{implant_aligned} or nanoimplantation techniques, \cite{FIB_implant} which would provide more reliable collection efficiency enhancement through control of the NV center lateral position, particularly with preferential dipole orientation alignment in (111)-oriented substrates. \cite{osterkamp_engineering_2019} This procedure could be extended to the fabrication of low-dimensional structures such as diamond nanosheets and nanowires, opening up a novel avenue for studying carrier\cite{lozovoi2021optical} and heat\cite{li2012thermal} transport in diamond.

\begin{acknowledgments}
This work was primarily supported by the Gordon and Betty Moore Foundation (grant GBMF12237, DOI 10.37807) and by the National Science Foundation under Grant No. OMA-2326767. Low-damage diamond etching was primarily supported by the U.S. Department of Energy, Office of Science, Office of Fusion Energy Sciences, under Award Number DE-SC0024537.

We thank Alex Abulnaga for help with the initial FDTD code we used in this work. The authors acknowledge the use of Princeton's Imaging and Analysis Center (IAC), which is partially supported by the Princeton Center for Complex Materials (PCCM), a National Science Foundation (NSF) Materials Research Science and Engineering Center (MRSEC; DMR-2011750) and John J. Schreiber for assistance in FIB and imaging. The devices in this work were fabricated at the Princeton Materials Institute Micro/Nanofabrication Center and Quantum Device Nanofabrication Laboratory - Princeton University. We also thank HQ graphene for their help and support building the micromanipulator transfer microscope.

A.C.P-T. developed the design and fabrication process, and performed measurements. A.L. carried out the spin and charge measurements. S.K. assisted with fabrication development and with T.B.M. conducted the modeling, photonic simulations, and PL measurements. M.L. performed the FIB cross-sectional analysis and assisted with fabrication. K.-H.C. and M.L. built the confocal microscope used for PL measurements. N.P.dL. supervised the project. A.C.P-T, A.L, S.K and N.P.dL wrote the manuscript.

\end{acknowledgments}

\textbf{Data availability:}
The data that support the findings of this study are available from the corresponding author upon reasonable request.

%%%%%%%%%%%%%%%%%%%%%%%%%%%%%%%%%%%%%%%%%%%%%%%%%%%%%%%%%%%%%%%%%%%%%%%%%%%%%%%%%%%%%%%%%%%%%%%%%%%%%%%%
%%%%%%%%%%%%%%%%%References
\bibliography{bib}

\begin{thebibliography}{48}%
\makeatletter
\providecommand \@ifxundefined [1]{%
 \@ifx{#1\undefined}
}%
\providecommand \@ifnum [1]{%
 \ifnum #1\expandafter \@firstoftwo
 \else \expandafter \@secondoftwo
 \fi
}%
\providecommand \@ifx [1]{%
 \ifx #1\expandafter \@firstoftwo
 \else \expandafter \@secondoftwo
 \fi
}%
\providecommand \natexlab [1]{#1}%
\providecommand \enquote  [1]{``#1''}%
\providecommand \bibnamefont  [1]{#1}%
\providecommand \bibfnamefont [1]{#1}%
\providecommand \citenamefont [1]{#1}%
\providecommand \href@noop [0]{\@secondoftwo}%
\providecommand \href [0]{\begingroup \@sanitize@url \@href}%
\providecommand \@href[1]{\@@startlink{#1}\@@href}%
\providecommand \@@href[1]{\endgroup#1\@@endlink}%
\providecommand \@sanitize@url [0]{\catcode `\\12\catcode `\$12\catcode `\&12\catcode `\#12\catcode `\^12\catcode `\_12\catcode `\%12\relax}%
\providecommand \@@startlink[1]{}%
\providecommand \@@endlink[0]{}%
\providecommand \url  [0]{\begingroup\@sanitize@url \@url }%
\providecommand \@url [1]{\endgroup\@href {#1}{\urlprefix }}%
\providecommand \urlprefix  [0]{URL }%
\providecommand \Eprint [0]{\href }%
\providecommand \doibase [0]{https://doi.org/}%
\providecommand \selectlanguage [0]{\@gobble}%
\providecommand \bibinfo  [0]{\@secondoftwo}%
\providecommand \bibfield  [0]{\@secondoftwo}%
\providecommand \translation [1]{[#1]}%
\providecommand \BibitemOpen [0]{}%
\providecommand \bibitemStop [0]{}%
\providecommand \bibitemNoStop [0]{.\EOS\space}%
\providecommand \EOS [0]{\spacefactor3000\relax}%
\providecommand \BibitemShut  [1]{\csname bibitem#1\endcsname}%
\let\auto@bib@innerbib\@empty
%</preamble>
\bibitem [{\citenamefont {Casola}\ \emph {et~al.}(2018)\citenamefont {Casola}, \citenamefont {Van Der~Sar},\ and\ \citenamefont {Yacoby}}]{casola2018probing}%
  \BibitemOpen
  \bibfield  {author} {\bibinfo {author} {\bibfnamefont {F.}~\bibnamefont {Casola}}, \bibinfo {author} {\bibfnamefont {T.}~\bibnamefont {Van Der~Sar}},\ and\ \bibinfo {author} {\bibfnamefont {A.}~\bibnamefont {Yacoby}},\ }\bibfield  {title} {\bibinfo {title} {Probing condensed matter physics with magnetometry based on nitrogen-vacancy centres in diamond},\ }\href@noop {} {\bibfield  {journal} {\bibinfo  {journal} {Nature Reviews Materials}\ }\textbf {\bibinfo {volume} {3}},\ \bibinfo {pages} {1} (\bibinfo {year} {2018})}\BibitemShut {NoStop}%
\bibitem [{\citenamefont {Aslam}\ \emph {et~al.}(2023)\citenamefont {Aslam}, \citenamefont {Zhou}, \citenamefont {Urbach}, \citenamefont {Turner}, \citenamefont {Walsworth}, \citenamefont {Lukin},\ and\ \citenamefont {Park}}]{aslam2023quantum}%
  \BibitemOpen
  \bibfield  {author} {\bibinfo {author} {\bibfnamefont {N.}~\bibnamefont {Aslam}}, \bibinfo {author} {\bibfnamefont {H.}~\bibnamefont {Zhou}}, \bibinfo {author} {\bibfnamefont {E.~K.}\ \bibnamefont {Urbach}}, \bibinfo {author} {\bibfnamefont {M.~J.}\ \bibnamefont {Turner}}, \bibinfo {author} {\bibfnamefont {R.~L.}\ \bibnamefont {Walsworth}}, \bibinfo {author} {\bibfnamefont {M.~D.}\ \bibnamefont {Lukin}},\ and\ \bibinfo {author} {\bibfnamefont {H.}~\bibnamefont {Park}},\ }\bibfield  {title} {\bibinfo {title} {Quantum sensors for biomedical applications},\ }\href@noop {} {\bibfield  {journal} {\bibinfo  {journal} {Nature Reviews Physics}\ }\textbf {\bibinfo {volume} {5}},\ \bibinfo {pages} {157} (\bibinfo {year} {2023})}\BibitemShut {NoStop}%
\bibitem [{\citenamefont {Rovny}\ \emph {et~al.}(2024)\citenamefont {Rovny} \emph {et~al.}}]{Rovny_Rev}%
  \BibitemOpen
  \bibfield  {author} {\bibinfo {author} {\bibfnamefont {J.}~\bibnamefont {Rovny}} \emph {et~al.},\ }\bibfield  {title} {\bibinfo {title} {New opportunities in condensed matter physics for nanoscale quantum sensors},\ }\href@noop {} {\bibfield  {journal} {\bibinfo  {journal} {arXiv preprint arXiv:2403.13710}\ } (\bibinfo {year} {2024})}\BibitemShut {NoStop}%
\bibitem [{\citenamefont {Kolkowitz}\ \emph {et~al.}(2015)\citenamefont {Kolkowitz}, \citenamefont {Safira}, \citenamefont {High}, \citenamefont {Devlin}, \citenamefont {Choi}, \citenamefont {Unterreithmeier}, \citenamefont {Patterson}, \citenamefont {Zibrov}, \citenamefont {Manucharyan}, \citenamefont {Park},\ and\ \citenamefont {Lukin}}]{kolkowitz2015probing}%
  \BibitemOpen
  \bibfield  {author} {\bibinfo {author} {\bibfnamefont {S.}~\bibnamefont {Kolkowitz}}, \bibinfo {author} {\bibfnamefont {A.}~\bibnamefont {Safira}}, \bibinfo {author} {\bibfnamefont {A.~A.}\ \bibnamefont {High}}, \bibinfo {author} {\bibfnamefont {R.~C.}\ \bibnamefont {Devlin}}, \bibinfo {author} {\bibfnamefont {S.}~\bibnamefont {Choi}}, \bibinfo {author} {\bibfnamefont {Q.~P.}\ \bibnamefont {Unterreithmeier}}, \bibinfo {author} {\bibfnamefont {D.}~\bibnamefont {Patterson}}, \bibinfo {author} {\bibfnamefont {A.~S.}\ \bibnamefont {Zibrov}}, \bibinfo {author} {\bibfnamefont {V.~E.}\ \bibnamefont {Manucharyan}}, \bibinfo {author} {\bibfnamefont {H.}~\bibnamefont {Park}},\ and\ \bibinfo {author} {\bibfnamefont {M.~D.}\ \bibnamefont {Lukin}},\ }\bibfield  {title} {\bibinfo {title} {Probing johnson noise and ballistic transport in normal metals with a single-spin qubit},\ }\href {https://doi.org/10.1126/science.aaa4298} {\bibfield  {journal} {\bibinfo  {journal} {Science}\ }\textbf {\bibinfo {volume} {347}},\
  \bibinfo {pages} {1129} (\bibinfo {year} {2015})},\ \Eprint {https://arxiv.org/abs/https://www.science.org/doi/pdf/10.1126/science.aaa4298} {https://www.science.org/doi/pdf/10.1126/science.aaa4298} \BibitemShut {NoStop}%
\bibitem [{\citenamefont {Andersen}\ \emph {et~al.}(2019)\citenamefont {Andersen}, \citenamefont {Dwyer}, \citenamefont {Sanchez-Yamagishi}, \citenamefont {Rodriguez-Nieva}, \citenamefont {Agarwal}, \citenamefont {Watanabe}, \citenamefont {Taniguchi}, \citenamefont {Demler}, \citenamefont {Kim}, \citenamefont {Park},\ and\ \citenamefont {Lukin}}]{andersen2019electron}%
  \BibitemOpen
  \bibfield  {author} {\bibinfo {author} {\bibfnamefont {T.~I.}\ \bibnamefont {Andersen}}, \bibinfo {author} {\bibfnamefont {B.~L.}\ \bibnamefont {Dwyer}}, \bibinfo {author} {\bibfnamefont {J.~D.}\ \bibnamefont {Sanchez-Yamagishi}}, \bibinfo {author} {\bibfnamefont {J.~F.}\ \bibnamefont {Rodriguez-Nieva}}, \bibinfo {author} {\bibfnamefont {K.}~\bibnamefont {Agarwal}}, \bibinfo {author} {\bibfnamefont {K.}~\bibnamefont {Watanabe}}, \bibinfo {author} {\bibfnamefont {T.}~\bibnamefont {Taniguchi}}, \bibinfo {author} {\bibfnamefont {E.~A.}\ \bibnamefont {Demler}}, \bibinfo {author} {\bibfnamefont {P.}~\bibnamefont {Kim}}, \bibinfo {author} {\bibfnamefont {H.}~\bibnamefont {Park}},\ and\ \bibinfo {author} {\bibfnamefont {M.~D.}\ \bibnamefont {Lukin}},\ }\bibfield  {title} {\bibinfo {title} {Electron-phonon instability in graphene revealed by global and local noise probes},\ }\href {https://doi.org/10.1126/science.aaw2104} {\bibfield  {journal} {\bibinfo  {journal} {Science}\ }\textbf {\bibinfo {volume} {364}},\
  \bibinfo {pages} {154} (\bibinfo {year} {2019})},\ \bibinfo {note} {\_eprint: https://www.science.org/doi/pdf/10.1126/science.aaw2104}\BibitemShut {NoStop}%
\bibitem [{\citenamefont {Vool}\ \emph {et~al.}(2021)\citenamefont {Vool}, \citenamefont {Hamo}, \citenamefont {Varnavides}, \citenamefont {Wang}, \citenamefont {Zhou}, \citenamefont {Kumar}, \citenamefont {Dovzhenko}, \citenamefont {Qiu}, \citenamefont {Garcia}, \citenamefont {Pierce}, \citenamefont {Gooth}, \citenamefont {Anikeeva}, \citenamefont {Felser}, \citenamefont {Narang},\ and\ \citenamefont {Yacoby}}]{vool2021imaging}%
  \BibitemOpen
  \bibfield  {author} {\bibinfo {author} {\bibfnamefont {U.}~\bibnamefont {Vool}}, \bibinfo {author} {\bibfnamefont {A.}~\bibnamefont {Hamo}}, \bibinfo {author} {\bibfnamefont {G.}~\bibnamefont {Varnavides}}, \bibinfo {author} {\bibfnamefont {Y.}~\bibnamefont {Wang}}, \bibinfo {author} {\bibfnamefont {T.~X.}\ \bibnamefont {Zhou}}, \bibinfo {author} {\bibfnamefont {N.}~\bibnamefont {Kumar}}, \bibinfo {author} {\bibfnamefont {Y.}~\bibnamefont {Dovzhenko}}, \bibinfo {author} {\bibfnamefont {Z.}~\bibnamefont {Qiu}}, \bibinfo {author} {\bibfnamefont {C.~A.~C.}\ \bibnamefont {Garcia}}, \bibinfo {author} {\bibfnamefont {A.~T.}\ \bibnamefont {Pierce}}, \bibinfo {author} {\bibfnamefont {J.}~\bibnamefont {Gooth}}, \bibinfo {author} {\bibfnamefont {P.}~\bibnamefont {Anikeeva}}, \bibinfo {author} {\bibfnamefont {C.}~\bibnamefont {Felser}}, \bibinfo {author} {\bibfnamefont {P.}~\bibnamefont {Narang}},\ and\ \bibinfo {author} {\bibfnamefont {A.}~\bibnamefont {Yacoby}},\ }\bibfield  {title} {\bibinfo {title} {Imaging
  phonon-mediated hydrodynamic flow in {WTe2}},\ }\href {https://doi.org/10.1038/s41567-021-01341-w} {\bibfield  {journal} {\bibinfo  {journal} {Nature Physics}\ }\textbf {\bibinfo {volume} {17}},\ \bibinfo {pages} {1216} (\bibinfo {year} {2021})}\BibitemShut {NoStop}%
\bibitem [{\citenamefont {Palm}\ \emph {et~al.}(2024)\citenamefont {Palm}, \citenamefont {Ding}, \citenamefont {Huxter}, \citenamefont {Taniguchi}, \citenamefont {Watanabe},\ and\ \citenamefont {Degen}}]{palm2024observation}%
  \BibitemOpen
  \bibfield  {author} {\bibinfo {author} {\bibfnamefont {M.~L.}\ \bibnamefont {Palm}}, \bibinfo {author} {\bibfnamefont {C.}~\bibnamefont {Ding}}, \bibinfo {author} {\bibfnamefont {W.~S.}\ \bibnamefont {Huxter}}, \bibinfo {author} {\bibfnamefont {T.}~\bibnamefont {Taniguchi}}, \bibinfo {author} {\bibfnamefont {K.}~\bibnamefont {Watanabe}},\ and\ \bibinfo {author} {\bibfnamefont {C.~L.}\ \bibnamefont {Degen}},\ }\bibfield  {title} {\bibinfo {title} {Observation of current whirlpools in graphene at room temperature},\ }\href@noop {} {\bibfield  {journal} {\bibinfo  {journal} {Science}\ }\textbf {\bibinfo {volume} {384}},\ \bibinfo {pages} {465} (\bibinfo {year} {2024})}\BibitemShut {NoStop}%
\bibitem [{\citenamefont {Huxter}\ \emph {et~al.}(2023)\citenamefont {Huxter}, \citenamefont {Sarott}, \citenamefont {Trassin},\ and\ \citenamefont {Degen}}]{huxter2023imaging}%
  \BibitemOpen
  \bibfield  {author} {\bibinfo {author} {\bibfnamefont {W.~S.}\ \bibnamefont {Huxter}}, \bibinfo {author} {\bibfnamefont {M.~F.}\ \bibnamefont {Sarott}}, \bibinfo {author} {\bibfnamefont {M.}~\bibnamefont {Trassin}},\ and\ \bibinfo {author} {\bibfnamefont {C.~L.}\ \bibnamefont {Degen}},\ }\bibfield  {title} {\bibinfo {title} {Imaging ferroelectric domains with a single-spin scanning quantum sensor},\ }\href@noop {} {\bibfield  {journal} {\bibinfo  {journal} {Nature Physics}\ }\textbf {\bibinfo {volume} {19}},\ \bibinfo {pages} {644} (\bibinfo {year} {2023})}\BibitemShut {NoStop}%
\bibitem [{\citenamefont {Tan}\ \emph {et~al.}(2024)\citenamefont {Tan}, \citenamefont {Jani}, \citenamefont {Högen}, \citenamefont {Stefan}, \citenamefont {Castelnovo}, \citenamefont {Braund}, \citenamefont {Geim}, \citenamefont {Mechnich}, \citenamefont {Feuer}, \citenamefont {Knowles}, \citenamefont {Ariando}, \citenamefont {Radaelli},\ and\ \citenamefont {Atatüre}}]{tan2024revealing}%
  \BibitemOpen
  \bibfield  {author} {\bibinfo {author} {\bibfnamefont {A.~K.~C.}\ \bibnamefont {Tan}}, \bibinfo {author} {\bibfnamefont {H.}~\bibnamefont {Jani}}, \bibinfo {author} {\bibfnamefont {M.}~\bibnamefont {Högen}}, \bibinfo {author} {\bibfnamefont {L.}~\bibnamefont {Stefan}}, \bibinfo {author} {\bibfnamefont {C.}~\bibnamefont {Castelnovo}}, \bibinfo {author} {\bibfnamefont {D.}~\bibnamefont {Braund}}, \bibinfo {author} {\bibfnamefont {A.}~\bibnamefont {Geim}}, \bibinfo {author} {\bibfnamefont {A.}~\bibnamefont {Mechnich}}, \bibinfo {author} {\bibfnamefont {M.~S.~G.}\ \bibnamefont {Feuer}}, \bibinfo {author} {\bibfnamefont {H.~S.}\ \bibnamefont {Knowles}}, \bibinfo {author} {\bibfnamefont {A.}~\bibnamefont {Ariando}}, \bibinfo {author} {\bibfnamefont {P.~G.}\ \bibnamefont {Radaelli}},\ and\ \bibinfo {author} {\bibfnamefont {M.}~\bibnamefont {Atatüre}},\ }\bibfield  {title} {\bibinfo {title} {Revealing emergent magnetic charge in an antiferromagnet with diamond quantum magnetometry},\ }\href
  {https://doi.org/10.1038/s41563-023-01737-4} {\bibfield  {journal} {\bibinfo  {journal} {Nature Materials}\ }\textbf {\bibinfo {volume} {23}},\ \bibinfo {pages} {205} (\bibinfo {year} {2024})}\BibitemShut {NoStop}%
\bibitem [{\citenamefont {Tschudin}\ \emph {et~al.}(2024)\citenamefont {Tschudin}, \citenamefont {Broadway}, \citenamefont {Siegwolf}, \citenamefont {Schrader}, \citenamefont {Telford}, \citenamefont {Gross}, \citenamefont {Cox}, \citenamefont {Dubois}, \citenamefont {Chica}, \citenamefont {Rama-Eiroa}, \citenamefont {Santos}, \citenamefont {Poggio}, \citenamefont {Ziebel}, \citenamefont {Dean}, \citenamefont {Roy},\ and\ \citenamefont {Maletinsky}}]{tschudin2024imaging}%
  \BibitemOpen
  \bibfield  {author} {\bibinfo {author} {\bibfnamefont {M.~A.}\ \bibnamefont {Tschudin}}, \bibinfo {author} {\bibfnamefont {D.~A.}\ \bibnamefont {Broadway}}, \bibinfo {author} {\bibfnamefont {P.}~\bibnamefont {Siegwolf}}, \bibinfo {author} {\bibfnamefont {C.}~\bibnamefont {Schrader}}, \bibinfo {author} {\bibfnamefont {E.~J.}\ \bibnamefont {Telford}}, \bibinfo {author} {\bibfnamefont {B.}~\bibnamefont {Gross}}, \bibinfo {author} {\bibfnamefont {J.}~\bibnamefont {Cox}}, \bibinfo {author} {\bibfnamefont {A.~E.~E.}\ \bibnamefont {Dubois}}, \bibinfo {author} {\bibfnamefont {D.~G.}\ \bibnamefont {Chica}}, \bibinfo {author} {\bibfnamefont {R.}~\bibnamefont {Rama-Eiroa}}, \bibinfo {author} {\bibfnamefont {E.~J.~G.}\ \bibnamefont {Santos}}, \bibinfo {author} {\bibfnamefont {M.}~\bibnamefont {Poggio}}, \bibinfo {author} {\bibfnamefont {M.~E.}\ \bibnamefont {Ziebel}}, \bibinfo {author} {\bibfnamefont {C.~R.}\ \bibnamefont {Dean}}, \bibinfo {author} {\bibfnamefont {X.}~\bibnamefont {Roy}},\ and\ \bibinfo {author}
  {\bibfnamefont {P.}~\bibnamefont {Maletinsky}},\ }\bibfield  {title} {\bibinfo {title} {Imaging nanomagnetism and magnetic phase transitions in atomically thin crsbr},\ }\href {https://doi.org/10.1038/s41467-024-49717-9} {\bibfield  {journal} {\bibinfo  {journal} {Nature Communications}\ }\textbf {\bibinfo {volume} {15}},\ \bibinfo {pages} {6005} (\bibinfo {year} {2024})}\BibitemShut {NoStop}%
\bibitem [{\citenamefont {Ziffer}\ \emph {et~al.}(2024)\citenamefont {Ziffer}, \citenamefont {Machado}, \citenamefont {Ursprung}, \citenamefont {Lozovoi}, \citenamefont {Tazi}, \citenamefont {Yuan}, \citenamefont {Ziebel}, \citenamefont {Delord}, \citenamefont {Zeng}, \citenamefont {Telford} \emph {et~al.}}]{ziffer2024quantum}%
  \BibitemOpen
  \bibfield  {author} {\bibinfo {author} {\bibfnamefont {M.~E.}\ \bibnamefont {Ziffer}}, \bibinfo {author} {\bibfnamefont {F.}~\bibnamefont {Machado}}, \bibinfo {author} {\bibfnamefont {B.}~\bibnamefont {Ursprung}}, \bibinfo {author} {\bibfnamefont {A.}~\bibnamefont {Lozovoi}}, \bibinfo {author} {\bibfnamefont {A.~B.}\ \bibnamefont {Tazi}}, \bibinfo {author} {\bibfnamefont {Z.}~\bibnamefont {Yuan}}, \bibinfo {author} {\bibfnamefont {M.~E.}\ \bibnamefont {Ziebel}}, \bibinfo {author} {\bibfnamefont {T.}~\bibnamefont {Delord}}, \bibinfo {author} {\bibfnamefont {N.}~\bibnamefont {Zeng}}, \bibinfo {author} {\bibfnamefont {E.}~\bibnamefont {Telford}}, \emph {et~al.},\ }\bibfield  {title} {\bibinfo {title} {Quantum noise spectroscopy of criticality in an atomically thin magnet},\ }\href@noop {} {\bibfield  {journal} {\bibinfo  {journal} {arXiv preprint arXiv:2407.05614}\ } (\bibinfo {year} {2024})}\BibitemShut {NoStop}%
\bibitem [{\citenamefont {Liu}\ \emph {et~al.}(2025)\citenamefont {Liu}, \citenamefont {Gong}, \citenamefont {Kim}, \citenamefont {Diessel}, \citenamefont {Xu}, \citenamefont {Rehfuss}, \citenamefont {Du}, \citenamefont {He}, \citenamefont {Singh}, \citenamefont {Eo} \emph {et~al.}}]{liu2025quantum}%
  \BibitemOpen
  \bibfield  {author} {\bibinfo {author} {\bibfnamefont {Z.}~\bibnamefont {Liu}}, \bibinfo {author} {\bibfnamefont {R.}~\bibnamefont {Gong}}, \bibinfo {author} {\bibfnamefont {J.}~\bibnamefont {Kim}}, \bibinfo {author} {\bibfnamefont {O.~K.}\ \bibnamefont {Diessel}}, \bibinfo {author} {\bibfnamefont {Q.}~\bibnamefont {Xu}}, \bibinfo {author} {\bibfnamefont {Z.}~\bibnamefont {Rehfuss}}, \bibinfo {author} {\bibfnamefont {X.}~\bibnamefont {Du}}, \bibinfo {author} {\bibfnamefont {G.}~\bibnamefont {He}}, \bibinfo {author} {\bibfnamefont {A.}~\bibnamefont {Singh}}, \bibinfo {author} {\bibfnamefont {Y.~S.}\ \bibnamefont {Eo}}, \emph {et~al.},\ }\bibfield  {title} {\bibinfo {title} {Quantum noise spectroscopy of superconducting dynamics in thin film $bi_2 sr_2 cacu_2 o_{8+\delta}$},\ }\href@noop {} {\bibfield  {journal} {\bibinfo  {journal} {arXiv preprint arXiv:2502.04439}\ } (\bibinfo {year} {2025})}\BibitemShut {NoStop}%
\bibitem [{\citenamefont {Maletinsky}\ \emph {et~al.}(2012)\citenamefont {Maletinsky}, \citenamefont {Hong}, \citenamefont {Grinolds}, \citenamefont {Hausmann}, \citenamefont {Lukin}, \citenamefont {Walsworth}, \citenamefont {Loncar},\ and\ \citenamefont {Yacoby}}]{maletinsky2012robust}%
  \BibitemOpen
  \bibfield  {author} {\bibinfo {author} {\bibfnamefont {P.}~\bibnamefont {Maletinsky}}, \bibinfo {author} {\bibfnamefont {S.}~\bibnamefont {Hong}}, \bibinfo {author} {\bibfnamefont {M.~S.}\ \bibnamefont {Grinolds}}, \bibinfo {author} {\bibfnamefont {B.}~\bibnamefont {Hausmann}}, \bibinfo {author} {\bibfnamefont {M.~D.}\ \bibnamefont {Lukin}}, \bibinfo {author} {\bibfnamefont {R.~L.}\ \bibnamefont {Walsworth}}, \bibinfo {author} {\bibfnamefont {M.}~\bibnamefont {Loncar}},\ and\ \bibinfo {author} {\bibfnamefont {A.}~\bibnamefont {Yacoby}},\ }\bibfield  {title} {\bibinfo {title} {A robust scanning diamond sensor for nanoscale imaging with single nitrogen-vacancy centres},\ }\href@noop {} {\bibfield  {journal} {\bibinfo  {journal} {Nature nanotechnology}\ }\textbf {\bibinfo {volume} {7}},\ \bibinfo {pages} {320} (\bibinfo {year} {2012})}\BibitemShut {NoStop}%
\bibitem [{\citenamefont {Scheidegger}\ \emph {et~al.}(2022)\citenamefont {Scheidegger}, \citenamefont {Diesch}, \citenamefont {Palm},\ and\ \citenamefont {Degen}}]{scheidegger2022scanning}%
  \BibitemOpen
  \bibfield  {author} {\bibinfo {author} {\bibfnamefont {P.~J.}\ \bibnamefont {Scheidegger}}, \bibinfo {author} {\bibfnamefont {S.}~\bibnamefont {Diesch}}, \bibinfo {author} {\bibfnamefont {M.~L.}\ \bibnamefont {Palm}},\ and\ \bibinfo {author} {\bibfnamefont {C.~L.}\ \bibnamefont {Degen}},\ }\bibfield  {title} {\bibinfo {title} {Scanning nitrogen-vacancy magnetometry down to 350 mk},\ }\href@noop {} {\bibfield  {journal} {\bibinfo  {journal} {Applied Physics Letters}\ }\textbf {\bibinfo {volume} {120}} (\bibinfo {year} {2022})}\BibitemShut {NoStop}%
\bibitem [{\citenamefont {Bluvstein}\ \emph {et~al.}(2019)\citenamefont {Bluvstein}, \citenamefont {Zhang},\ and\ \citenamefont {Jayich}}]{bluvstein2019identifying}%
  \BibitemOpen
  \bibfield  {author} {\bibinfo {author} {\bibfnamefont {D.}~\bibnamefont {Bluvstein}}, \bibinfo {author} {\bibfnamefont {Z.}~\bibnamefont {Zhang}},\ and\ \bibinfo {author} {\bibfnamefont {A.~C.~B.}\ \bibnamefont {Jayich}},\ }\bibfield  {title} {\bibinfo {title} {Identifying and mitigating charge instabilities in shallow diamond nitrogen-vacancy centers},\ }\href@noop {} {\bibfield  {journal} {\bibinfo  {journal} {Physical review letters}\ }\textbf {\bibinfo {volume} {122}},\ \bibinfo {pages} {076101} (\bibinfo {year} {2019})}\BibitemShut {NoStop}%
\bibitem [{\citenamefont {Yuan}\ \emph {et~al.}(2020)\citenamefont {Yuan}, \citenamefont {Fitzpatrick}, \citenamefont {Rodgers}, \citenamefont {Sangtawesin}, \citenamefont {Srinivasan},\ and\ \citenamefont {De~Leon}}]{yuan2020charge}%
  \BibitemOpen
  \bibfield  {author} {\bibinfo {author} {\bibfnamefont {Z.}~\bibnamefont {Yuan}}, \bibinfo {author} {\bibfnamefont {M.}~\bibnamefont {Fitzpatrick}}, \bibinfo {author} {\bibfnamefont {L.~V.}\ \bibnamefont {Rodgers}}, \bibinfo {author} {\bibfnamefont {S.}~\bibnamefont {Sangtawesin}}, \bibinfo {author} {\bibfnamefont {S.}~\bibnamefont {Srinivasan}},\ and\ \bibinfo {author} {\bibfnamefont {N.~P.}\ \bibnamefont {De~Leon}},\ }\bibfield  {title} {\bibinfo {title} {Charge state dynamics and optically detected electron spin resonance contrast of shallow nitrogen-vacancy centers in diamond},\ }\href@noop {} {\bibfield  {journal} {\bibinfo  {journal} {Physical Review Research}\ }\textbf {\bibinfo {volume} {2}},\ \bibinfo {pages} {033263} (\bibinfo {year} {2020})}\BibitemShut {NoStop}%
\bibitem [{\citenamefont {Leibold}\ \emph {et~al.}(2025)\citenamefont {Leibold}, \citenamefont {Todenhagen}, \citenamefont {Althammer}, \citenamefont {Khera}, \citenamefont {Neu}, \citenamefont {Brandt}, \citenamefont {Huebl},\ and\ \citenamefont {Bucher}}]{leibold2025influence}%
  \BibitemOpen
  \bibfield  {author} {\bibinfo {author} {\bibfnamefont {J.~P.}\ \bibnamefont {Leibold}}, \bibinfo {author} {\bibfnamefont {L.~M.}\ \bibnamefont {Todenhagen}}, \bibinfo {author} {\bibfnamefont {M.}~\bibnamefont {Althammer}}, \bibinfo {author} {\bibfnamefont {N.}~\bibnamefont {Khera}}, \bibinfo {author} {\bibfnamefont {E.}~\bibnamefont {Neu}}, \bibinfo {author} {\bibfnamefont {M.~S.}\ \bibnamefont {Brandt}}, \bibinfo {author} {\bibfnamefont {H.}~\bibnamefont {Huebl}},\ and\ \bibinfo {author} {\bibfnamefont {D.~B.}\ \bibnamefont {Bucher}},\ }\bibfield  {title} {\bibinfo {title} {Influence of platinum thin films on the photophysical and quantum properties of near-surface nv centers},\ }\href@noop {} {\bibfield  {journal} {\bibinfo  {journal} {arXiv preprint arXiv:2510.11721}\ }\textbf {\bibinfo {volume} {1}} (\bibinfo {year} {2025})}\BibitemShut {NoStop}%
\bibitem [{\citenamefont {Sangtawesin}\ \emph {et~al.}(2019)\citenamefont {Sangtawesin}, \citenamefont {Dwyer}, \citenamefont {Srinivasan}, \citenamefont {Allred}, \citenamefont {Rodgers}, \citenamefont {De~Greve}, \citenamefont {Stacey}, \citenamefont {Dontschuk}, \citenamefont {O'Donnell}, \citenamefont {Hu}, \citenamefont {Evans}, \citenamefont {Jaye}, \citenamefont {Fischer}, \citenamefont {Markham}, \citenamefont {Twitchen}, \citenamefont {Park}, \citenamefont {Lukin},\ and\ \citenamefont {de~Leon}}]{Sangtawesin_SurfNoise}%
  \BibitemOpen
  \bibfield  {author} {\bibinfo {author} {\bibfnamefont {S.}~\bibnamefont {Sangtawesin}}, \bibinfo {author} {\bibfnamefont {B.~L.}\ \bibnamefont {Dwyer}}, \bibinfo {author} {\bibfnamefont {S.}~\bibnamefont {Srinivasan}}, \bibinfo {author} {\bibfnamefont {J.~J.}\ \bibnamefont {Allred}}, \bibinfo {author} {\bibfnamefont {L.~V.~H.}\ \bibnamefont {Rodgers}}, \bibinfo {author} {\bibfnamefont {K.}~\bibnamefont {De~Greve}}, \bibinfo {author} {\bibfnamefont {A.}~\bibnamefont {Stacey}}, \bibinfo {author} {\bibfnamefont {N.}~\bibnamefont {Dontschuk}}, \bibinfo {author} {\bibfnamefont {K.~M.}\ \bibnamefont {O'Donnell}}, \bibinfo {author} {\bibfnamefont {D.}~\bibnamefont {Hu}}, \bibinfo {author} {\bibfnamefont {D.~A.}\ \bibnamefont {Evans}}, \bibinfo {author} {\bibfnamefont {C.}~\bibnamefont {Jaye}}, \bibinfo {author} {\bibfnamefont {D.~A.}\ \bibnamefont {Fischer}}, \bibinfo {author} {\bibfnamefont {M.~L.}\ \bibnamefont {Markham}}, \bibinfo {author} {\bibfnamefont {D.~J.}\ \bibnamefont {Twitchen}}, \bibinfo {author}
  {\bibfnamefont {H.}~\bibnamefont {Park}}, \bibinfo {author} {\bibfnamefont {M.~D.}\ \bibnamefont {Lukin}},\ and\ \bibinfo {author} {\bibfnamefont {N.~P.}\ \bibnamefont {de~Leon}},\ }\bibfield  {title} {\bibinfo {title} {Origins of diamond surface noise probed by correlating single-spin measurements with surface spectroscopy},\ }\href {https://doi.org/10.1103/PhysRevX.9.031052} {\bibfield  {journal} {\bibinfo  {journal} {Phys. Rev. X}\ }\textbf {\bibinfo {volume} {9}},\ \bibinfo {pages} {031052} (\bibinfo {year} {2019})}\BibitemShut {NoStop}%
\bibitem [{\citenamefont {Corazza}\ \emph {et~al.}(2025)\citenamefont {Corazza}, \citenamefont {Ruffieux}, \citenamefont {Zhu}, \citenamefont {Jaramillo~Concha}, \citenamefont {Fontana}, \citenamefont {Galland}, \citenamefont {Warburton},\ and\ \citenamefont {Maletinsky}}]{corazza2025homogeneous}%
  \BibitemOpen
  \bibfield  {author} {\bibinfo {author} {\bibfnamefont {A.}~\bibnamefont {Corazza}}, \bibinfo {author} {\bibfnamefont {S.}~\bibnamefont {Ruffieux}}, \bibinfo {author} {\bibfnamefont {Y.}~\bibnamefont {Zhu}}, \bibinfo {author} {\bibfnamefont {C.~A.}\ \bibnamefont {Jaramillo~Concha}}, \bibinfo {author} {\bibfnamefont {Y.}~\bibnamefont {Fontana}}, \bibinfo {author} {\bibfnamefont {C.}~\bibnamefont {Galland}}, \bibinfo {author} {\bibfnamefont {R.~J.}\ \bibnamefont {Warburton}},\ and\ \bibinfo {author} {\bibfnamefont {P.}~\bibnamefont {Maletinsky}},\ }\bibfield  {title} {\bibinfo {title} {Homogeneous free-standing nanostructures from bulk diamond over millimeter scales for quantum technologies},\ }\href@noop {} {\bibfield  {journal} {\bibinfo  {journal} {Nano Letters}\ }\textbf {\bibinfo {volume} {25}} (\bibinfo {year} {2025})}\BibitemShut {NoStop}%
\bibitem [{\citenamefont {Momenzadeh}\ \emph {et~al.}(2015)\citenamefont {Momenzadeh}, \citenamefont {Stöhr}, \citenamefont {Oliveira}, \citenamefont {Brunner}, \citenamefont {Denisenko}, \citenamefont {Yang}, \citenamefont {Reinhard},\ and\ \citenamefont {Wrachtrup}}]{Wrachtrup_pillar}%
  \BibitemOpen
  \bibfield  {author} {\bibinfo {author} {\bibfnamefont {S.~A.}\ \bibnamefont {Momenzadeh}}, \bibinfo {author} {\bibfnamefont {R.~J.}\ \bibnamefont {Stöhr}}, \bibinfo {author} {\bibfnamefont {F.~F.~d.}\ \bibnamefont {Oliveira}}, \bibinfo {author} {\bibfnamefont {A.}~\bibnamefont {Brunner}}, \bibinfo {author} {\bibfnamefont {A.}~\bibnamefont {Denisenko}}, \bibinfo {author} {\bibfnamefont {S.}~\bibnamefont {Yang}}, \bibinfo {author} {\bibfnamefont {F.}~\bibnamefont {Reinhard}},\ and\ \bibinfo {author} {\bibfnamefont {J.}~\bibnamefont {Wrachtrup}},\ }\bibfield  {title} {\bibinfo {title} {{Nanoengineered Diamond Waveguide as a Robust Bright Platform for Nanomagnetometry Using Shallow Nitrogen Vacancy Centers}},\ }\href {https://doi.org/10.1021/nl503326t} {\bibfield  {journal} {\bibinfo  {journal} {Nano Letters}\ }\textbf {\bibinfo {volume} {15}},\ \bibinfo {pages} {165} (\bibinfo {year} {2015})},\ \Eprint {https://arxiv.org/abs/1409.0027} {1409.0027} \BibitemShut {NoStop}%
\bibitem [{\citenamefont {Thiel}\ \emph {et~al.}(2016)\citenamefont {Thiel}, \citenamefont {Rohner}, \citenamefont {Ganzhorn}, \citenamefont {Appel}, \citenamefont {Neu}, \citenamefont {Müller}, \citenamefont {Kleiner}, \citenamefont {Koelle},\ and\ \citenamefont {Maletinsky}}]{Maletinsky_AFMCryo}%
  \BibitemOpen
  \bibfield  {author} {\bibinfo {author} {\bibfnamefont {L.}~\bibnamefont {Thiel}}, \bibinfo {author} {\bibfnamefont {D.}~\bibnamefont {Rohner}}, \bibinfo {author} {\bibfnamefont {M.}~\bibnamefont {Ganzhorn}}, \bibinfo {author} {\bibfnamefont {P.}~\bibnamefont {Appel}}, \bibinfo {author} {\bibfnamefont {E.}~\bibnamefont {Neu}}, \bibinfo {author} {\bibfnamefont {B.}~\bibnamefont {Müller}}, \bibinfo {author} {\bibfnamefont {R.}~\bibnamefont {Kleiner}}, \bibinfo {author} {\bibfnamefont {D.}~\bibnamefont {Koelle}},\ and\ \bibinfo {author} {\bibfnamefont {P.}~\bibnamefont {Maletinsky}},\ }\bibfield  {title} {\bibinfo {title} {{Quantitative nanoscale vortex imaging using a cryogenic quantum magnetometer}},\ }\href {https://doi.org/10.1038/nnano.2016.63} {\bibfield  {journal} {\bibinfo  {journal} {Nature Nanotechnology}\ }\textbf {\bibinfo {volume} {11}},\ \bibinfo {pages} {677} (\bibinfo {year} {2016})},\ \Eprint {https://arxiv.org/abs/1511.02873} {1511.02873} \BibitemShut {NoStop}%
\bibitem [{\citenamefont {Burek}\ \emph {et~al.}(2012)\citenamefont {Burek}, \citenamefont {Leon}, \citenamefont {Shields}, \citenamefont {Hausmann}, \citenamefont {Chu}, \citenamefont {Quan}, \citenamefont {Zibrov}, \citenamefont {Park}, \citenamefont {Lukin},\ and\ \citenamefont {Lončar}}]{Loncar_beams_fab}%
  \BibitemOpen
  \bibfield  {author} {\bibinfo {author} {\bibfnamefont {M.~J.}\ \bibnamefont {Burek}}, \bibinfo {author} {\bibfnamefont {N.~P.~d.}\ \bibnamefont {Leon}}, \bibinfo {author} {\bibfnamefont {B.~J.}\ \bibnamefont {Shields}}, \bibinfo {author} {\bibfnamefont {B.~J.~M.}\ \bibnamefont {Hausmann}}, \bibinfo {author} {\bibfnamefont {Y.}~\bibnamefont {Chu}}, \bibinfo {author} {\bibfnamefont {Q.}~\bibnamefont {Quan}}, \bibinfo {author} {\bibfnamefont {A.~S.}\ \bibnamefont {Zibrov}}, \bibinfo {author} {\bibfnamefont {H.}~\bibnamefont {Park}}, \bibinfo {author} {\bibfnamefont {M.~D.}\ \bibnamefont {Lukin}},\ and\ \bibinfo {author} {\bibfnamefont {M.}~\bibnamefont {Lončar}},\ }\bibfield  {title} {\bibinfo {title} {{Free-Standing Mechanical and Photonic Nanostructures in Single-Crystal Diamond}},\ }\href {https://doi.org/10.1021/nl302541e} {\bibfield  {journal} {\bibinfo  {journal} {Nano Letters}\ }\textbf {\bibinfo {volume} {12}},\ \bibinfo {pages} {6084} (\bibinfo {year} {2012})}\BibitemShut {NoStop}%
\bibitem [{\citenamefont {Du}\ \emph {et~al.}(2017)\citenamefont {Du}, \citenamefont {van~der Sar}, \citenamefont {Zhou}, \citenamefont {Upadhyaya}, \citenamefont {Casola}, \citenamefont {Zhang}, \citenamefont {Onbasli}, \citenamefont {Ross}, \citenamefont {Walsworth}, \citenamefont {Tserkovnyak},\ and\ \citenamefont {Yacoby}}]{Yacoby_tribeams_chempot}%
  \BibitemOpen
  \bibfield  {author} {\bibinfo {author} {\bibfnamefont {C.}~\bibnamefont {Du}}, \bibinfo {author} {\bibfnamefont {T.}~\bibnamefont {van~der Sar}}, \bibinfo {author} {\bibfnamefont {T.~X.}\ \bibnamefont {Zhou}}, \bibinfo {author} {\bibfnamefont {P.}~\bibnamefont {Upadhyaya}}, \bibinfo {author} {\bibfnamefont {F.}~\bibnamefont {Casola}}, \bibinfo {author} {\bibfnamefont {H.}~\bibnamefont {Zhang}}, \bibinfo {author} {\bibfnamefont {M.~C.}\ \bibnamefont {Onbasli}}, \bibinfo {author} {\bibfnamefont {C.~A.}\ \bibnamefont {Ross}}, \bibinfo {author} {\bibfnamefont {R.~L.}\ \bibnamefont {Walsworth}}, \bibinfo {author} {\bibfnamefont {Y.}~\bibnamefont {Tserkovnyak}},\ and\ \bibinfo {author} {\bibfnamefont {A.}~\bibnamefont {Yacoby}},\ }\bibfield  {title} {\bibinfo {title} {Control and local measurement of the spin chemical potential in a magnetic insulator},\ }\href {https://doi.org/10.1126/science.aak9611} {\bibfield  {journal} {\bibinfo  {journal} {Science}\ }\textbf {\bibinfo {volume} {357}},\ \bibinfo {pages}
  {195} (\bibinfo {year} {2017})},\ \Eprint {https://arxiv.org/abs/https://www.science.org/doi/pdf/10.1126/science.aak9611} {https://www.science.org/doi/pdf/10.1126/science.aak9611} \BibitemShut {NoStop}%
\bibitem [{\citenamefont {Wang}\ \emph {et~al.}(2022{\natexlab{a}})\citenamefont {Wang}, \citenamefont {Zhang}, \citenamefont {McLaughlin}, \citenamefont {Flebus}, \citenamefont {Huang}, \citenamefont {Xiao}, \citenamefont {Liu}, \citenamefont {Wu}, \citenamefont {Fullerton}, \citenamefont {Tserkovnyak},\ and\ \citenamefont {Du}}]{Du_tribeams_Fe}%
  \BibitemOpen
  \bibfield  {author} {\bibinfo {author} {\bibfnamefont {H.}~\bibnamefont {Wang}}, \bibinfo {author} {\bibfnamefont {S.}~\bibnamefont {Zhang}}, \bibinfo {author} {\bibfnamefont {N.~J.}\ \bibnamefont {McLaughlin}}, \bibinfo {author} {\bibfnamefont {B.}~\bibnamefont {Flebus}}, \bibinfo {author} {\bibfnamefont {M.}~\bibnamefont {Huang}}, \bibinfo {author} {\bibfnamefont {Y.}~\bibnamefont {Xiao}}, \bibinfo {author} {\bibfnamefont {C.}~\bibnamefont {Liu}}, \bibinfo {author} {\bibfnamefont {M.}~\bibnamefont {Wu}}, \bibinfo {author} {\bibfnamefont {E.~E.}\ \bibnamefont {Fullerton}}, \bibinfo {author} {\bibfnamefont {Y.}~\bibnamefont {Tserkovnyak}},\ and\ \bibinfo {author} {\bibfnamefont {C.~R.}\ \bibnamefont {Du}},\ }\bibfield  {title} {\bibinfo {title} {Noninvasive measurements of spin transport properties of an antiferromagnetic insulator},\ }\href {https://doi.org/10.1126/sciadv.abg8562} {\bibfield  {journal} {\bibinfo  {journal} {Science Advances}\ }\textbf {\bibinfo {volume} {8}},\ \bibinfo {pages} {eabg8562}
  (\bibinfo {year} {2022}{\natexlab{a}})},\ \Eprint {https://arxiv.org/abs/https://www.science.org/doi/pdf/10.1126/sciadv.abg8562} {https://www.science.org/doi/pdf/10.1126/sciadv.abg8562} \BibitemShut {NoStop}%
\bibitem [{\citenamefont {Guo}\ \emph {et~al.}(2021)\citenamefont {Guo}, \citenamefont {Delegan}, \citenamefont {Karsch}, \citenamefont {Li}, \citenamefont {Liu}, \citenamefont {Shreiner}, \citenamefont {Butcher}, \citenamefont {Awschalom}, \citenamefont {Heremans},\ and\ \citenamefont {High}}]{Membrane_2_High}%
  \BibitemOpen
  \bibfield  {author} {\bibinfo {author} {\bibfnamefont {X.}~\bibnamefont {Guo}}, \bibinfo {author} {\bibfnamefont {N.}~\bibnamefont {Delegan}}, \bibinfo {author} {\bibfnamefont {J.~C.}\ \bibnamefont {Karsch}}, \bibinfo {author} {\bibfnamefont {Z.}~\bibnamefont {Li}}, \bibinfo {author} {\bibfnamefont {T.}~\bibnamefont {Liu}}, \bibinfo {author} {\bibfnamefont {R.}~\bibnamefont {Shreiner}}, \bibinfo {author} {\bibfnamefont {A.}~\bibnamefont {Butcher}}, \bibinfo {author} {\bibfnamefont {D.~D.}\ \bibnamefont {Awschalom}}, \bibinfo {author} {\bibfnamefont {F.~J.}\ \bibnamefont {Heremans}},\ and\ \bibinfo {author} {\bibfnamefont {A.~A.}\ \bibnamefont {High}},\ }\bibfield  {title} {\bibinfo {title} {{Tunable and Transferable Diamond Membranes for Integrated Quantum Technologies}},\ }\href {https://doi.org/10.1021/acs.nanolett.1c03703} {\bibfield  {journal} {\bibinfo  {journal} {Nano Letters}\ }\textbf {\bibinfo {volume} {21}},\ \bibinfo {pages} {10392} (\bibinfo {year} {2021})},\ \Eprint
  {https://arxiv.org/abs/2109.11507} {2109.11507} \BibitemShut {NoStop}%
\bibitem [{\citenamefont {Guo}\ \emph {et~al.}(2024)\citenamefont {Guo}, \citenamefont {Xie}, \citenamefont {Addhya}, \citenamefont {Linder}, \citenamefont {Zvi}, \citenamefont {Wang}, \citenamefont {Yu}, \citenamefont {Deshmukh}, \citenamefont {Liu}, \citenamefont {Hammock}, \citenamefont {Li}, \citenamefont {DeVault}, \citenamefont {Butcher}, \citenamefont {Esser-Kahn}, \citenamefont {Awschalom}, \citenamefont {Delegan}, \citenamefont {Maurer}, \citenamefont {Heremans},\ and\ \citenamefont {High}}]{guo_direct-bonded_2024}%
  \BibitemOpen
  \bibfield  {author} {\bibinfo {author} {\bibfnamefont {X.}~\bibnamefont {Guo}}, \bibinfo {author} {\bibfnamefont {M.}~\bibnamefont {Xie}}, \bibinfo {author} {\bibfnamefont {A.}~\bibnamefont {Addhya}}, \bibinfo {author} {\bibfnamefont {A.}~\bibnamefont {Linder}}, \bibinfo {author} {\bibfnamefont {U.}~\bibnamefont {Zvi}}, \bibinfo {author} {\bibfnamefont {S.}~\bibnamefont {Wang}}, \bibinfo {author} {\bibfnamefont {X.}~\bibnamefont {Yu}}, \bibinfo {author} {\bibfnamefont {T.~D.}\ \bibnamefont {Deshmukh}}, \bibinfo {author} {\bibfnamefont {Y.}~\bibnamefont {Liu}}, \bibinfo {author} {\bibfnamefont {I.~N.}\ \bibnamefont {Hammock}}, \bibinfo {author} {\bibfnamefont {Z.}~\bibnamefont {Li}}, \bibinfo {author} {\bibfnamefont {C.~T.}\ \bibnamefont {DeVault}}, \bibinfo {author} {\bibfnamefont {A.}~\bibnamefont {Butcher}}, \bibinfo {author} {\bibfnamefont {A.~P.}\ \bibnamefont {Esser-Kahn}}, \bibinfo {author} {\bibfnamefont {D.~D.}\ \bibnamefont {Awschalom}}, \bibinfo {author} {\bibfnamefont {N.}~\bibnamefont
  {Delegan}}, \bibinfo {author} {\bibfnamefont {P.~C.}\ \bibnamefont {Maurer}}, \bibinfo {author} {\bibfnamefont {F.~J.}\ \bibnamefont {Heremans}},\ and\ \bibinfo {author} {\bibfnamefont {A.~A.}\ \bibnamefont {High}},\ }\bibfield  {title} {\bibinfo {title} {Direct-bonded diamond membranes for heterogeneous quantum and electronic technologies},\ }\href {https://doi.org/10.1038/s41467-024-53150-3} {\bibfield  {journal} {\bibinfo  {journal} {Nature Communications}\ }\textbf {\bibinfo {volume} {15}},\ \bibinfo {pages} {8788} (\bibinfo {year} {2024})}\BibitemShut {NoStop}%
\bibitem [{\citenamefont {Ding}\ \emph {et~al.}(2024)\citenamefont {Ding}, \citenamefont {Haas}, \citenamefont {Guo}, \citenamefont {Kuruma}, \citenamefont {Jin}, \citenamefont {Li}, \citenamefont {Awschalom}, \citenamefont {Delegan}, \citenamefont {Heremans}, \citenamefont {High},\ and\ \citenamefont {Loncar}}]{ding_high-q_2024}%
  \BibitemOpen
  \bibfield  {author} {\bibinfo {author} {\bibfnamefont {S.~W.}\ \bibnamefont {Ding}}, \bibinfo {author} {\bibfnamefont {M.}~\bibnamefont {Haas}}, \bibinfo {author} {\bibfnamefont {X.}~\bibnamefont {Guo}}, \bibinfo {author} {\bibfnamefont {K.}~\bibnamefont {Kuruma}}, \bibinfo {author} {\bibfnamefont {C.}~\bibnamefont {Jin}}, \bibinfo {author} {\bibfnamefont {Z.}~\bibnamefont {Li}}, \bibinfo {author} {\bibfnamefont {D.~D.}\ \bibnamefont {Awschalom}}, \bibinfo {author} {\bibfnamefont {N.}~\bibnamefont {Delegan}}, \bibinfo {author} {\bibfnamefont {F.~J.}\ \bibnamefont {Heremans}}, \bibinfo {author} {\bibfnamefont {A.~A.}\ \bibnamefont {High}},\ and\ \bibinfo {author} {\bibfnamefont {M.}~\bibnamefont {Loncar}},\ }\bibfield  {title} {\bibinfo {title} {High-{Q} cavity interface for color centers in thin film diamond},\ }\href {https://doi.org/10.1038/s41467-024-50667-5} {\bibfield  {journal} {\bibinfo  {journal} {Nature Communications}\ }\textbf {\bibinfo {volume} {15}},\ \bibinfo {pages} {6358} (\bibinfo {year}
  {2024})}\BibitemShut {NoStop}%
\bibitem [{\citenamefont {Shaw}\ \emph {et~al.}(1994)\citenamefont {Shaw}, \citenamefont {Zhang},\ and\ \citenamefont {MacDonald}}]{SCREAM}%
  \BibitemOpen
  \bibfield  {author} {\bibinfo {author} {\bibfnamefont {K.~A.}\ \bibnamefont {Shaw}}, \bibinfo {author} {\bibfnamefont {Z.}~\bibnamefont {Zhang}},\ and\ \bibinfo {author} {\bibfnamefont {N.~C.}\ \bibnamefont {MacDonald}},\ }\bibfield  {title} {\bibinfo {title} {Scream i: A single mask, single-crystal silicon, reactive ion etching process for microelectromechanical structures},\ }\href {https://doi.org/https://doi.org/10.1016/0924-4247(94)85031-3} {\bibfield  {journal} {\bibinfo  {journal} {Sensors and Actuators A: Physical}\ }\textbf {\bibinfo {volume} {40}},\ \bibinfo {pages} {63} (\bibinfo {year} {1994})}\BibitemShut {NoStop}%
\bibitem [{\citenamefont {Khanaliloo}\ \emph {et~al.}(2015{\natexlab{a}})\citenamefont {Khanaliloo}, \citenamefont {Mitchell}, \citenamefont {Hryciw},\ and\ \citenamefont {Barclay}}]{Khanaliloo_highQ}%
  \BibitemOpen
  \bibfield  {author} {\bibinfo {author} {\bibfnamefont {B.}~\bibnamefont {Khanaliloo}}, \bibinfo {author} {\bibfnamefont {M.}~\bibnamefont {Mitchell}}, \bibinfo {author} {\bibfnamefont {A.~C.}\ \bibnamefont {Hryciw}},\ and\ \bibinfo {author} {\bibfnamefont {P.~E.}\ \bibnamefont {Barclay}},\ }\bibfield  {title} {\bibinfo {title} {{High‑Q/V Monolithic Diamond Microdisks Fabricated with Quasi-isotropic Etching}},\ }\href {https://doi.org/10.1021/acs.nanolett.5b01346} {\bibfield  {journal} {\bibinfo  {journal} {Nano Letters}\ }\textbf {\bibinfo {volume} {15}},\ \bibinfo {pages} {5131} (\bibinfo {year} {2015}{\natexlab{a}})}\BibitemShut {NoStop}%
\bibitem [{\citenamefont {Khanaliloo}\ \emph {et~al.}(2015{\natexlab{b}})\citenamefont {Khanaliloo}, \citenamefont {Jayakumar}, \citenamefont {Hryciw}, \citenamefont {Lake}, \citenamefont {Kaviani},\ and\ \citenamefont {Barclay}}]{Barclay_waveguide}%
  \BibitemOpen
  \bibfield  {author} {\bibinfo {author} {\bibfnamefont {B.}~\bibnamefont {Khanaliloo}}, \bibinfo {author} {\bibfnamefont {H.}~\bibnamefont {Jayakumar}}, \bibinfo {author} {\bibfnamefont {A.~C.}\ \bibnamefont {Hryciw}}, \bibinfo {author} {\bibfnamefont {D.~P.}\ \bibnamefont {Lake}}, \bibinfo {author} {\bibfnamefont {H.}~\bibnamefont {Kaviani}},\ and\ \bibinfo {author} {\bibfnamefont {P.~E.}\ \bibnamefont {Barclay}},\ }\bibfield  {title} {\bibinfo {title} {Single-crystal diamond nanobeam waveguide optomechanics},\ }\href {https://doi.org/10.1103/PhysRevX.5.041051} {\bibfield  {journal} {\bibinfo  {journal} {Phys. Rev. X}\ }\textbf {\bibinfo {volume} {5}},\ \bibinfo {pages} {041051} (\bibinfo {year} {2015}{\natexlab{b}})}\BibitemShut {NoStop}%
\bibitem [{\citenamefont {Xie}\ \emph {et~al.}(2018)\citenamefont {Xie}, \citenamefont {Zhou}, \citenamefont {Stöhr},\ and\ \citenamefont {Yacoby}}]{Yacoby_isodeepdive}%
  \BibitemOpen
  \bibfield  {author} {\bibinfo {author} {\bibfnamefont {L.}~\bibnamefont {Xie}}, \bibinfo {author} {\bibfnamefont {T.~X.}\ \bibnamefont {Zhou}}, \bibinfo {author} {\bibfnamefont {R.~J.}\ \bibnamefont {Stöhr}},\ and\ \bibinfo {author} {\bibfnamefont {A.}~\bibnamefont {Yacoby}},\ }\bibfield  {title} {\bibinfo {title} {Crystallographic orientation dependent reactive ion etching in single crystal diamond},\ }\href {https://doi.org/https://doi.org/10.1002/adma.201705501} {\bibfield  {journal} {\bibinfo  {journal} {Advanced Materials}\ }\textbf {\bibinfo {volume} {30}},\ \bibinfo {pages} {1705501} (\bibinfo {year} {2018})},\ \Eprint {https://arxiv.org/abs/https://advanced.onlinelibrary.wiley.com/doi/pdf/10.1002/adma.201705501} {https://advanced.onlinelibrary.wiley.com/doi/pdf/10.1002/adma.201705501} \BibitemShut {NoStop}%
\bibitem [{\citenamefont {Mitchell}\ \emph {et~al.}(2019)\citenamefont {Mitchell}, \citenamefont {Lake},\ and\ \citenamefont {Barclay}}]{Barclay_300000}%
  \BibitemOpen
  \bibfield  {author} {\bibinfo {author} {\bibfnamefont {M.}~\bibnamefont {Mitchell}}, \bibinfo {author} {\bibfnamefont {D.~P.}\ \bibnamefont {Lake}},\ and\ \bibinfo {author} {\bibfnamefont {P.~E.}\ \bibnamefont {Barclay}},\ }\bibfield  {title} {\bibinfo {title} {Realizing q 300 000 in diamond microdisks for optomechanics via etch optimization},\ }\href {https://doi.org/10.1063/1.5053122} {\bibfield  {journal} {\bibinfo  {journal} {APL Photonics}\ }\textbf {\bibinfo {volume} {4}},\ \bibinfo {pages} {016101} (\bibinfo {year} {2019})}\BibitemShut {NoStop}%
\bibitem [{\citenamefont {Li}\ \emph {et~al.}(2023)\citenamefont {Li}, \citenamefont {Gerritsma}, \citenamefont {Kurdi}, \citenamefont {Codreanu}, \citenamefont {Gr{\"o}blacher}, \citenamefont {Hanson}, \citenamefont {Norte},\ and\ \citenamefont {van~der Sar}}]{TeonoFibreCoupled}%
  \BibitemOpen
  \bibfield  {author} {\bibinfo {author} {\bibfnamefont {Y.}~\bibnamefont {Li}}, \bibinfo {author} {\bibfnamefont {F.~A.}\ \bibnamefont {Gerritsma}}, \bibinfo {author} {\bibfnamefont {S.}~\bibnamefont {Kurdi}}, \bibinfo {author} {\bibfnamefont {N.}~\bibnamefont {Codreanu}}, \bibinfo {author} {\bibfnamefont {S.}~\bibnamefont {Gr{\"o}blacher}}, \bibinfo {author} {\bibfnamefont {R.}~\bibnamefont {Hanson}}, \bibinfo {author} {\bibfnamefont {R.}~\bibnamefont {Norte}},\ and\ \bibinfo {author} {\bibfnamefont {T.}~\bibnamefont {van~der Sar}},\ }\bibfield  {title} {\bibinfo {title} {A fiber-coupled scanning magnetometer with nitrogen-vacancy spins in a diamond nanobeam},\ }\href {https://doi.org/10.1021/acsphotonics.3c00259} {\bibfield  {journal} {\bibinfo  {journal} {ACS Photonics}\ }\textbf {\bibinfo {volume} {10}},\ \bibinfo {pages} {1859} (\bibinfo {year} {2023})},\ \Eprint {https://arxiv.org/abs/https://doi.org/10.1021/acsphotonics.3c00259} {https://doi.org/10.1021/acsphotonics.3c00259} \BibitemShut {NoStop}%
\bibitem [{\citenamefont {Pasini}\ \emph {et~al.}(2024)\citenamefont {Pasini}, \citenamefont {Codreanu}, \citenamefont {Turan}, \citenamefont {Riera~Moral}, \citenamefont {Primavera}, \citenamefont {De~Santis}, \citenamefont {Beukers}, \citenamefont {Brevoord}, \citenamefont {Waas}, \citenamefont {Borregaard},\ and\ \citenamefont {Hanson}}]{hanson_waveguide}%
  \BibitemOpen
  \bibfield  {author} {\bibinfo {author} {\bibfnamefont {M.}~\bibnamefont {Pasini}}, \bibinfo {author} {\bibfnamefont {N.}~\bibnamefont {Codreanu}}, \bibinfo {author} {\bibfnamefont {T.}~\bibnamefont {Turan}}, \bibinfo {author} {\bibfnamefont {A.}~\bibnamefont {Riera~Moral}}, \bibinfo {author} {\bibfnamefont {C.~F.}\ \bibnamefont {Primavera}}, \bibinfo {author} {\bibfnamefont {L.}~\bibnamefont {De~Santis}}, \bibinfo {author} {\bibfnamefont {H.~K.~C.}\ \bibnamefont {Beukers}}, \bibinfo {author} {\bibfnamefont {J.~M.}\ \bibnamefont {Brevoord}}, \bibinfo {author} {\bibfnamefont {C.}~\bibnamefont {Waas}}, \bibinfo {author} {\bibfnamefont {J.}~\bibnamefont {Borregaard}},\ and\ \bibinfo {author} {\bibfnamefont {R.}~\bibnamefont {Hanson}},\ }\bibfield  {title} {\bibinfo {title} {Nonlinear quantum photonics with a tin-vacancy center coupled to a one-dimensional diamond waveguide},\ }\href {https://doi.org/10.1103/PhysRevLett.133.023603} {\bibfield  {journal} {\bibinfo  {journal} {Phys. Rev. Lett.}\ }\textbf {\bibinfo
  {volume} {133}},\ \bibinfo {pages} {023603} (\bibinfo {year} {2024})}\BibitemShut {NoStop}%
\bibitem [{\citenamefont {Brevoord}\ \emph {et~al.}(2025)\citenamefont {Brevoord}, \citenamefont {Wienhoven}, \citenamefont {Codreanu}, \citenamefont {Ishiguro}, \citenamefont {van Leeuwen}, \citenamefont {Iuliano}, \citenamefont {De~Santis}, \citenamefont {Waas}, \citenamefont {Beukers}, \citenamefont {Turan}, \citenamefont {Errando-Herranz}, \citenamefont {Kawaguchi},\ and\ \citenamefont {Hanson}}]{brevoord_large-range_2025}%
  \BibitemOpen
  \bibfield  {author} {\bibinfo {author} {\bibfnamefont {J.~M.}\ \bibnamefont {Brevoord}}, \bibinfo {author} {\bibfnamefont {L.~G.~C.}\ \bibnamefont {Wienhoven}}, \bibinfo {author} {\bibfnamefont {N.}~\bibnamefont {Codreanu}}, \bibinfo {author} {\bibfnamefont {T.}~\bibnamefont {Ishiguro}}, \bibinfo {author} {\bibfnamefont {E.}~\bibnamefont {van Leeuwen}}, \bibinfo {author} {\bibfnamefont {M.}~\bibnamefont {Iuliano}}, \bibinfo {author} {\bibfnamefont {L.}~\bibnamefont {De~Santis}}, \bibinfo {author} {\bibfnamefont {C.}~\bibnamefont {Waas}}, \bibinfo {author} {\bibfnamefont {H.~K.~C.}\ \bibnamefont {Beukers}}, \bibinfo {author} {\bibfnamefont {T.}~\bibnamefont {Turan}}, \bibinfo {author} {\bibfnamefont {C.}~\bibnamefont {Errando-Herranz}}, \bibinfo {author} {\bibfnamefont {K.}~\bibnamefont {Kawaguchi}},\ and\ \bibinfo {author} {\bibfnamefont {R.}~\bibnamefont {Hanson}},\ }\bibfield  {title} {\bibinfo {title} {Large-range tuning and stabilization of the optical transition of diamond tin-vacancy centers by in
  situ strain control},\ }\href {https://doi.org/10.1063/5.0251211} {\bibfield  {journal} {\bibinfo  {journal} {Applied Physics Letters}\ }\textbf {\bibinfo {volume} {126}},\ \bibinfo {pages} {174001} (\bibinfo {year} {2025})},\ \bibinfo {note} {\_eprint: https://pubs.aip.org/aip/apl/article-pdf/doi/10.1063/5.0251211/20503511/174001\_1\_5.0251211.pdf}\BibitemShut {NoStop}%
\bibitem [{\citenamefont {Mouradian}\ \emph {et~al.}(2017)\citenamefont {Mouradian}, \citenamefont {Wan}, \citenamefont {Schröder},\ and\ \citenamefont {Englund}}]{Englund_phc}%
  \BibitemOpen
  \bibfield  {author} {\bibinfo {author} {\bibfnamefont {S.}~\bibnamefont {Mouradian}}, \bibinfo {author} {\bibfnamefont {N.~H.}\ \bibnamefont {Wan}}, \bibinfo {author} {\bibfnamefont {T.}~\bibnamefont {Schröder}},\ and\ \bibinfo {author} {\bibfnamefont {D.}~\bibnamefont {Englund}},\ }\bibfield  {title} {\bibinfo {title} {Rectangular photonic crystal nanobeam cavities in bulk diamond},\ }\href {https://doi.org/10.1063/1.4992118} {\bibfield  {journal} {\bibinfo  {journal} {Applied Physics Letters}\ }\textbf {\bibinfo {volume} {111}},\ \bibinfo {pages} {021103} (\bibinfo {year} {2017})}\BibitemShut {NoStop}%
\bibitem [{\citenamefont {Banerjee}\ \emph {et~al.}(2018)\citenamefont {Banerjee}, \citenamefont {Bernoulli}, \citenamefont {Zhang}, \citenamefont {Yuen}, \citenamefont {Liu}, \citenamefont {Dong}, \citenamefont {Ding}, \citenamefont {Lu}, \citenamefont {Dao}, \citenamefont {Zhang}, \citenamefont {Lu},\ and\ \citenamefont {Suresh}}]{Suresh_mechFlex}%
  \BibitemOpen
  \bibfield  {author} {\bibinfo {author} {\bibfnamefont {A.}~\bibnamefont {Banerjee}}, \bibinfo {author} {\bibfnamefont {D.}~\bibnamefont {Bernoulli}}, \bibinfo {author} {\bibfnamefont {H.}~\bibnamefont {Zhang}}, \bibinfo {author} {\bibfnamefont {M.-F.}\ \bibnamefont {Yuen}}, \bibinfo {author} {\bibfnamefont {J.}~\bibnamefont {Liu}}, \bibinfo {author} {\bibfnamefont {J.}~\bibnamefont {Dong}}, \bibinfo {author} {\bibfnamefont {F.}~\bibnamefont {Ding}}, \bibinfo {author} {\bibfnamefont {J.}~\bibnamefont {Lu}}, \bibinfo {author} {\bibfnamefont {M.}~\bibnamefont {Dao}}, \bibinfo {author} {\bibfnamefont {W.}~\bibnamefont {Zhang}}, \bibinfo {author} {\bibfnamefont {Y.}~\bibnamefont {Lu}},\ and\ \bibinfo {author} {\bibfnamefont {S.}~\bibnamefont {Suresh}},\ }\bibfield  {title} {\bibinfo {title} {Ultralarge elastic deformation of nanoscale diamond},\ }\href {https://doi.org/10.1126/science.aar4165} {\bibfield  {journal} {\bibinfo  {journal} {Science}\ }\textbf {\bibinfo {volume} {360}},\ \bibinfo {pages} {300}
  (\bibinfo {year} {2018})},\ \Eprint {https://arxiv.org/abs/https://www.science.org/doi/pdf/10.1126/science.aar4165} {https://www.science.org/doi/pdf/10.1126/science.aar4165} \BibitemShut {NoStop}%
\bibitem [{\citenamefont {Pham}\ \emph {et~al.}(2016)\citenamefont {Pham}, \citenamefont {DeVience}, \citenamefont {Casola}, \citenamefont {Lovchinsky}, \citenamefont {Sushkov}, \citenamefont {Bersin}, \citenamefont {Lee}, \citenamefont {Urbach}, \citenamefont {Cappellaro}, \citenamefont {Park}, \citenamefont {Yacoby}, \citenamefont {Lukin},\ and\ \citenamefont {Walsworth}}]{pham2016nmr}%
  \BibitemOpen
  \bibfield  {author} {\bibinfo {author} {\bibfnamefont {L.~M.}\ \bibnamefont {Pham}}, \bibinfo {author} {\bibfnamefont {S.~J.}\ \bibnamefont {DeVience}}, \bibinfo {author} {\bibfnamefont {F.}~\bibnamefont {Casola}}, \bibinfo {author} {\bibfnamefont {I.}~\bibnamefont {Lovchinsky}}, \bibinfo {author} {\bibfnamefont {A.~O.}\ \bibnamefont {Sushkov}}, \bibinfo {author} {\bibfnamefont {E.}~\bibnamefont {Bersin}}, \bibinfo {author} {\bibfnamefont {J.}~\bibnamefont {Lee}}, \bibinfo {author} {\bibfnamefont {E.}~\bibnamefont {Urbach}}, \bibinfo {author} {\bibfnamefont {P.}~\bibnamefont {Cappellaro}}, \bibinfo {author} {\bibfnamefont {H.}~\bibnamefont {Park}}, \bibinfo {author} {\bibfnamefont {A.}~\bibnamefont {Yacoby}}, \bibinfo {author} {\bibfnamefont {M.}~\bibnamefont {Lukin}},\ and\ \bibinfo {author} {\bibfnamefont {R.~L.}\ \bibnamefont {Walsworth}},\ }\bibfield  {title} {\bibinfo {title} {Nmr technique for determining the depth of shallow nitrogen-vacancy centers in diamond},\ }\href
  {https://doi.org/10.1103/PhysRevB.93.045425} {\bibfield  {journal} {\bibinfo  {journal} {Phys. Rev. B}\ }\textbf {\bibinfo {volume} {93}},\ \bibinfo {pages} {045425} (\bibinfo {year} {2016})}\BibitemShut {NoStop}%
\bibitem [{\citenamefont {Myers}\ \emph {et~al.}(2014)\citenamefont {Myers}, \citenamefont {Das}, \citenamefont {Dartiailh}, \citenamefont {Ohno}, \citenamefont {Awschalom},\ and\ \citenamefont {Bleszynski~Jayich}}]{myers2014probing}%
  \BibitemOpen
  \bibfield  {author} {\bibinfo {author} {\bibfnamefont {B.~A.}\ \bibnamefont {Myers}}, \bibinfo {author} {\bibfnamefont {A.}~\bibnamefont {Das}}, \bibinfo {author} {\bibfnamefont {M.}~\bibnamefont {Dartiailh}}, \bibinfo {author} {\bibfnamefont {K.}~\bibnamefont {Ohno}}, \bibinfo {author} {\bibfnamefont {D.~D.}\ \bibnamefont {Awschalom}},\ and\ \bibinfo {author} {\bibfnamefont {A.}~\bibnamefont {Bleszynski~Jayich}},\ }\bibfield  {title} {\bibinfo {title} {Probing surface noise with depth-calibrated spins in diamond},\ }\href@noop {} {\bibfield  {journal} {\bibinfo  {journal} {Physical Review Letters}\ }\textbf {\bibinfo {volume} {113}},\ \bibinfo {pages} {027602} (\bibinfo {year} {2014})}\BibitemShut {NoStop}%
\bibitem [{\citenamefont {Bar-Gill}\ \emph {et~al.}(2012)\citenamefont {Bar-Gill}, \citenamefont {Pham}, \citenamefont {Belthangady}, \citenamefont {Le~Sage}, \citenamefont {Cappellaro}, \citenamefont {Maze}, \citenamefont {Lukin}, \citenamefont {Yacoby},\ and\ \citenamefont {Walsworth}}]{bar2012suppression}%
  \BibitemOpen
  \bibfield  {author} {\bibinfo {author} {\bibfnamefont {N.}~\bibnamefont {Bar-Gill}}, \bibinfo {author} {\bibfnamefont {L.~M.}\ \bibnamefont {Pham}}, \bibinfo {author} {\bibfnamefont {C.}~\bibnamefont {Belthangady}}, \bibinfo {author} {\bibfnamefont {D.}~\bibnamefont {Le~Sage}}, \bibinfo {author} {\bibfnamefont {P.}~\bibnamefont {Cappellaro}}, \bibinfo {author} {\bibfnamefont {J.}~\bibnamefont {Maze}}, \bibinfo {author} {\bibfnamefont {M.~D.}\ \bibnamefont {Lukin}}, \bibinfo {author} {\bibfnamefont {A.}~\bibnamefont {Yacoby}},\ and\ \bibinfo {author} {\bibfnamefont {R.}~\bibnamefont {Walsworth}},\ }\bibfield  {title} {\bibinfo {title} {Suppression of spin-bath dynamics for improved coherence of multi-spin-qubit systems},\ }\href@noop {} {\bibfield  {journal} {\bibinfo  {journal} {Nature communications}\ }\textbf {\bibinfo {volume} {3}},\ \bibinfo {pages} {858} (\bibinfo {year} {2012})}\BibitemShut {NoStop}%
\bibitem [{\citenamefont {Romach}\ \emph {et~al.}(2015)\citenamefont {Romach}, \citenamefont {M\"uller}, \citenamefont {Unden}, \citenamefont {Rogers}, \citenamefont {Isoda}, \citenamefont {Itoh}, \citenamefont {Markham}, \citenamefont {Stacey}, \citenamefont {Meijer}, \citenamefont {Pezzagna}, \citenamefont {Naydenov}, \citenamefont {McGuinness}, \citenamefont {Bar-Gill},\ and\ \citenamefont {Jelezko}}]{romach2015spectroscopy}%
  \BibitemOpen
  \bibfield  {author} {\bibinfo {author} {\bibfnamefont {Y.}~\bibnamefont {Romach}}, \bibinfo {author} {\bibfnamefont {C.}~\bibnamefont {M\"uller}}, \bibinfo {author} {\bibfnamefont {T.}~\bibnamefont {Unden}}, \bibinfo {author} {\bibfnamefont {L.~J.}\ \bibnamefont {Rogers}}, \bibinfo {author} {\bibfnamefont {T.}~\bibnamefont {Isoda}}, \bibinfo {author} {\bibfnamefont {K.~M.}\ \bibnamefont {Itoh}}, \bibinfo {author} {\bibfnamefont {M.}~\bibnamefont {Markham}}, \bibinfo {author} {\bibfnamefont {A.}~\bibnamefont {Stacey}}, \bibinfo {author} {\bibfnamefont {J.}~\bibnamefont {Meijer}}, \bibinfo {author} {\bibfnamefont {S.}~\bibnamefont {Pezzagna}}, \bibinfo {author} {\bibfnamefont {B.}~\bibnamefont {Naydenov}}, \bibinfo {author} {\bibfnamefont {L.~P.}\ \bibnamefont {McGuinness}}, \bibinfo {author} {\bibfnamefont {N.}~\bibnamefont {Bar-Gill}},\ and\ \bibinfo {author} {\bibfnamefont {F.}~\bibnamefont {Jelezko}},\ }\bibfield  {title} {\bibinfo {title} {Spectroscopy of surface-induced noise using shallow spins in
  diamond},\ }\href {https://doi.org/10.1103/PhysRevLett.114.017601} {\bibfield  {journal} {\bibinfo  {journal} {Phys. Rev. Lett.}\ }\textbf {\bibinfo {volume} {114}},\ \bibinfo {pages} {017601} (\bibinfo {year} {2015})}\BibitemShut {NoStop}%
\bibitem [{\citenamefont {Dhomkar}\ \emph {et~al.}(2018)\citenamefont {Dhomkar}, \citenamefont {Jayakumar}, \citenamefont {Zangara},\ and\ \citenamefont {Meriles}}]{dhomkar2018charge}%
  \BibitemOpen
  \bibfield  {author} {\bibinfo {author} {\bibfnamefont {S.}~\bibnamefont {Dhomkar}}, \bibinfo {author} {\bibfnamefont {H.}~\bibnamefont {Jayakumar}}, \bibinfo {author} {\bibfnamefont {P.~R.}\ \bibnamefont {Zangara}},\ and\ \bibinfo {author} {\bibfnamefont {C.~A.}\ \bibnamefont {Meriles}},\ }\bibfield  {title} {\bibinfo {title} {Charge dynamics in near-surface, variable-density ensembles of nitrogen-vacancy centers in diamond},\ }\href@noop {} {\bibfield  {journal} {\bibinfo  {journal} {Nano letters}\ }\textbf {\bibinfo {volume} {18}},\ \bibinfo {pages} {4046} (\bibinfo {year} {2018})}\BibitemShut {NoStop}%
\bibitem [{\citenamefont {Appel}\ \emph {et~al.}(2016)\citenamefont {Appel}, \citenamefont {Neu}, \citenamefont {Ganzhorn}, \citenamefont {Barfuss}, \citenamefont {Batzer}, \citenamefont {Gratz}, \citenamefont {Tschöpe},\ and\ \citenamefont {Maletinsky}}]{Maletinsky_tran}%
  \BibitemOpen
  \bibfield  {author} {\bibinfo {author} {\bibfnamefont {P.}~\bibnamefont {Appel}}, \bibinfo {author} {\bibfnamefont {E.}~\bibnamefont {Neu}}, \bibinfo {author} {\bibfnamefont {M.}~\bibnamefont {Ganzhorn}}, \bibinfo {author} {\bibfnamefont {A.}~\bibnamefont {Barfuss}}, \bibinfo {author} {\bibfnamefont {M.}~\bibnamefont {Batzer}}, \bibinfo {author} {\bibfnamefont {M.}~\bibnamefont {Gratz}}, \bibinfo {author} {\bibfnamefont {A.}~\bibnamefont {Tschöpe}},\ and\ \bibinfo {author} {\bibfnamefont {P.}~\bibnamefont {Maletinsky}},\ }\bibfield  {title} {\bibinfo {title} {Fabrication of all diamond scanning probes for nanoscale magnetometry},\ }\href {https://doi.org/10.1063/1.4952953} {\bibfield  {journal} {\bibinfo  {journal} {Review of Scientific Instruments}\ }\textbf {\bibinfo {volume} {87}},\ \bibinfo {pages} {063703} (\bibinfo {year} {2016})}\BibitemShut {NoStop}%
\bibitem [{\citenamefont {Wang}\ \emph {et~al.}(2022{\natexlab{b}})\citenamefont {Wang}, \citenamefont {Sun}, \citenamefont {Ye}, \citenamefont {Yu}, \citenamefont {Liu}, \citenamefont {Zhou}, \citenamefont {Wang}, \citenamefont {Shi}, \citenamefont {Wang},\ and\ \citenamefont {Du}}]{implant_aligned}%
  \BibitemOpen
  \bibfield  {author} {\bibinfo {author} {\bibfnamefont {M.}~\bibnamefont {Wang}}, \bibinfo {author} {\bibfnamefont {H.}~\bibnamefont {Sun}}, \bibinfo {author} {\bibfnamefont {X.}~\bibnamefont {Ye}}, \bibinfo {author} {\bibfnamefont {P.}~\bibnamefont {Yu}}, \bibinfo {author} {\bibfnamefont {H.}~\bibnamefont {Liu}}, \bibinfo {author} {\bibfnamefont {J.}~\bibnamefont {Zhou}}, \bibinfo {author} {\bibfnamefont {P.}~\bibnamefont {Wang}}, \bibinfo {author} {\bibfnamefont {F.}~\bibnamefont {Shi}}, \bibinfo {author} {\bibfnamefont {Y.}~\bibnamefont {Wang}},\ and\ \bibinfo {author} {\bibfnamefont {J.}~\bibnamefont {Du}},\ }\bibfield  {title} {\bibinfo {title} {Self-aligned patterning technique for fabricating high-performance diamond sensor arrays with nanoscale precision},\ }\href {https://doi.org/10.1126/sciadv.abn9573} {\bibfield  {journal} {\bibinfo  {journal} {Science Advances}\ }\textbf {\bibinfo {volume} {8}},\ \bibinfo {pages} {eabn9573} (\bibinfo {year} {2022}{\natexlab{b}})},\ \Eprint
  {https://arxiv.org/abs/https://www.science.org/doi/pdf/10.1126/sciadv.abn9573} {https://www.science.org/doi/pdf/10.1126/sciadv.abn9573} \BibitemShut {NoStop}%
\bibitem [{\citenamefont {Chandrasekaran}\ \emph {et~al.}(2023)\citenamefont {Chandrasekaran}, \citenamefont {Titze}, \citenamefont {Flores}, \citenamefont {Campbell}, \citenamefont {Henshaw}, \citenamefont {Jones}, \citenamefont {Bielejec},\ and\ \citenamefont {Htoon}}]{FIB_implant}%
  \BibitemOpen
  \bibfield  {author} {\bibinfo {author} {\bibfnamefont {V.}~\bibnamefont {Chandrasekaran}}, \bibinfo {author} {\bibfnamefont {M.}~\bibnamefont {Titze}}, \bibinfo {author} {\bibfnamefont {A.~R.}\ \bibnamefont {Flores}}, \bibinfo {author} {\bibfnamefont {D.}~\bibnamefont {Campbell}}, \bibinfo {author} {\bibfnamefont {J.}~\bibnamefont {Henshaw}}, \bibinfo {author} {\bibfnamefont {A.~C.}\ \bibnamefont {Jones}}, \bibinfo {author} {\bibfnamefont {E.~S.}\ \bibnamefont {Bielejec}},\ and\ \bibinfo {author} {\bibfnamefont {H.}~\bibnamefont {Htoon}},\ }\bibfield  {title} {\bibinfo {title} {High-yield deterministic focused ion beam implantation of quantum defects enabled by in situ photoluminescence feedback},\ }\href {https://doi.org/https://doi.org/10.1002/advs.202300190} {\bibfield  {journal} {\bibinfo  {journal} {Advanced Science}\ }\textbf {\bibinfo {volume} {10}},\ \bibinfo {pages} {2300190} (\bibinfo {year} {2023})},\ \Eprint
  {https://arxiv.org/abs/https://advanced.onlinelibrary.wiley.com/doi/pdf/10.1002/advs.202300190} {https://advanced.onlinelibrary.wiley.com/doi/pdf/10.1002/advs.202300190} \BibitemShut {NoStop}%
\bibitem [{\citenamefont {Osterkamp}\ \emph {et~al.}(2019)\citenamefont {Osterkamp}, \citenamefont {Mangold}, \citenamefont {Lang}, \citenamefont {Balasubramanian}, \citenamefont {Teraji}, \citenamefont {Naydenov},\ and\ \citenamefont {Jelezko}}]{osterkamp_engineering_2019}%
  \BibitemOpen
  \bibfield  {author} {\bibinfo {author} {\bibfnamefont {C.}~\bibnamefont {Osterkamp}}, \bibinfo {author} {\bibfnamefont {M.}~\bibnamefont {Mangold}}, \bibinfo {author} {\bibfnamefont {J.}~\bibnamefont {Lang}}, \bibinfo {author} {\bibfnamefont {P.}~\bibnamefont {Balasubramanian}}, \bibinfo {author} {\bibfnamefont {T.}~\bibnamefont {Teraji}}, \bibinfo {author} {\bibfnamefont {B.}~\bibnamefont {Naydenov}},\ and\ \bibinfo {author} {\bibfnamefont {F.}~\bibnamefont {Jelezko}},\ }\bibfield  {title} {\bibinfo {title} {Engineering preferentially-aligned nitrogen-vacancy centre ensembles in {CVD} grown diamond},\ }\href {https://doi.org/10.1038/s41598-019-42314-7} {\bibfield  {journal} {\bibinfo  {journal} {Scientific Reports}\ }\textbf {\bibinfo {volume} {9}},\ \bibinfo {pages} {5786} (\bibinfo {year} {2019})}\BibitemShut {NoStop}%
\bibitem [{\citenamefont {Lozovoi}\ \emph {et~al.}(2021)\citenamefont {Lozovoi}, \citenamefont {Jayakumar}, \citenamefont {Daw}, \citenamefont {Vizkelethy}, \citenamefont {Bielejec}, \citenamefont {Doherty}, \citenamefont {Flick},\ and\ \citenamefont {Meriles}}]{lozovoi2021optical}%
  \BibitemOpen
  \bibfield  {author} {\bibinfo {author} {\bibfnamefont {A.}~\bibnamefont {Lozovoi}}, \bibinfo {author} {\bibfnamefont {H.}~\bibnamefont {Jayakumar}}, \bibinfo {author} {\bibfnamefont {D.}~\bibnamefont {Daw}}, \bibinfo {author} {\bibfnamefont {G.}~\bibnamefont {Vizkelethy}}, \bibinfo {author} {\bibfnamefont {E.}~\bibnamefont {Bielejec}}, \bibinfo {author} {\bibfnamefont {M.~W.}\ \bibnamefont {Doherty}}, \bibinfo {author} {\bibfnamefont {J.}~\bibnamefont {Flick}},\ and\ \bibinfo {author} {\bibfnamefont {C.~A.}\ \bibnamefont {Meriles}},\ }\bibfield  {title} {\bibinfo {title} {Optical activation and detection of charge transport between individual colour centres in diamond},\ }\href@noop {} {\bibfield  {journal} {\bibinfo  {journal} {Nature Electronics}\ }\textbf {\bibinfo {volume} {4}},\ \bibinfo {pages} {717} (\bibinfo {year} {2021})}\BibitemShut {NoStop}%
\bibitem [{\citenamefont {Li}\ \emph {et~al.}(2012)\citenamefont {Li}, \citenamefont {Mingo}, \citenamefont {Lindsay}, \citenamefont {Broido}, \citenamefont {Stewart},\ and\ \citenamefont {Katcho}}]{li2012thermal}%
  \BibitemOpen
  \bibfield  {author} {\bibinfo {author} {\bibfnamefont {W.}~\bibnamefont {Li}}, \bibinfo {author} {\bibfnamefont {N.}~\bibnamefont {Mingo}}, \bibinfo {author} {\bibfnamefont {L.}~\bibnamefont {Lindsay}}, \bibinfo {author} {\bibfnamefont {D.~A.}\ \bibnamefont {Broido}}, \bibinfo {author} {\bibfnamefont {D.~A.}\ \bibnamefont {Stewart}},\ and\ \bibinfo {author} {\bibfnamefont {N.~A.}\ \bibnamefont {Katcho}},\ }\bibfield  {title} {\bibinfo {title} {Thermal conductivity of diamond nanowires from first principles},\ }\href@noop {} {\bibfield  {journal} {\bibinfo  {journal} {Physical Review B—Condensed Matter and Materials Physics}\ }\textbf {\bibinfo {volume} {85}},\ \bibinfo {pages} {195436} (\bibinfo {year} {2012})}\BibitemShut {NoStop}%
\end{thebibliography}%

\renewcommand{\thefigure}{S.\arabic{figure}}
\renewcommand\thesection{S.\arabic{section}}

\setcounter{section}{0}
\setcounter{figure}{0}
\newpage
\newpage
\newpage

\section{Fabrication Details}

\begin{figure}
    \centering
    \includegraphics[width=1\linewidth]{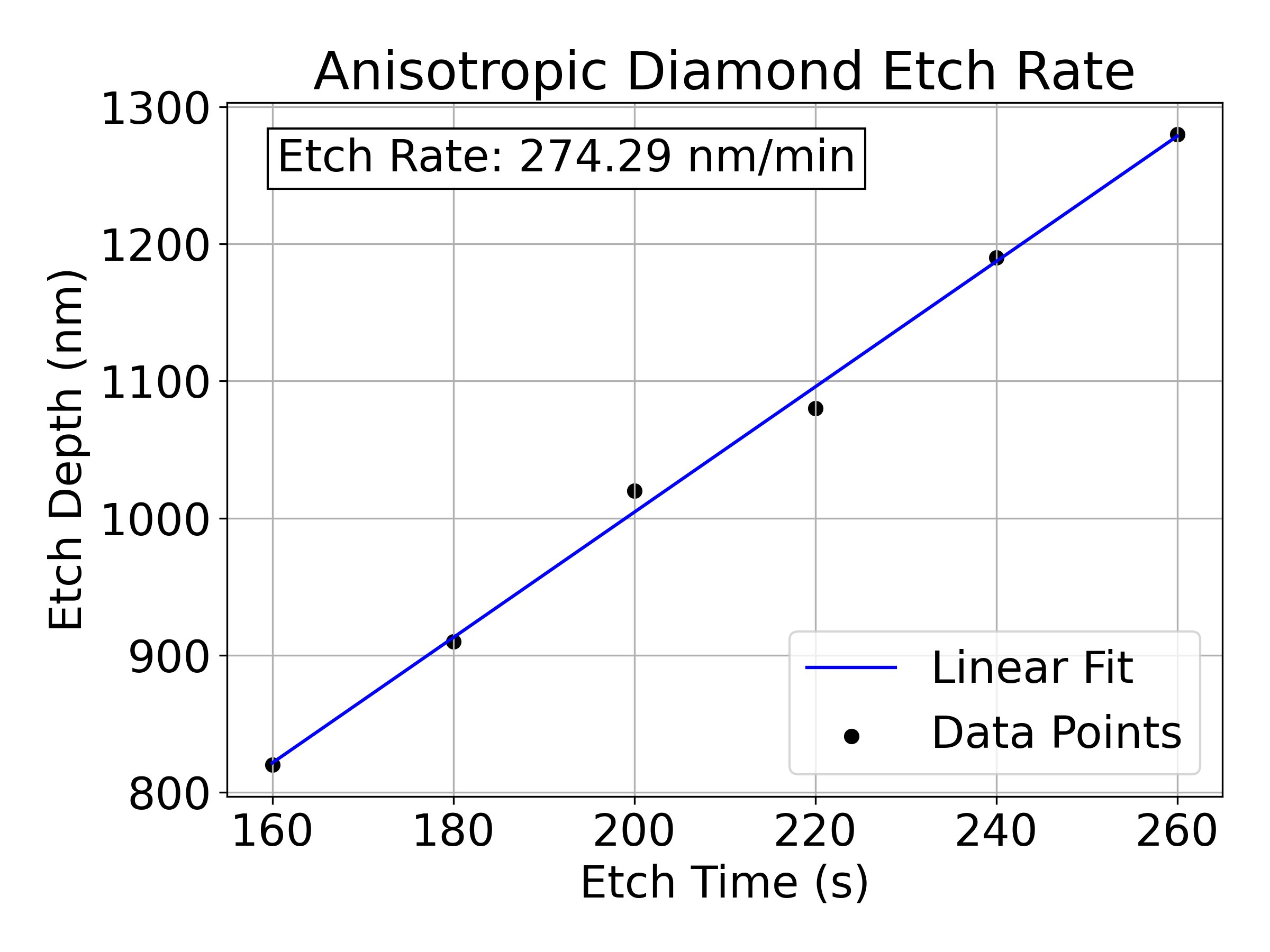
    }
    \caption{Anisotropic etch depth of diamond in oxygen ICP-RIE as a function of etch time. Blue line is a linear fit to the data that sets the etch rate of the recipe in main text}
    \label{fig:rate_1}
\end{figure}

The anisotropic etch process is calibrated over a number of test etches, Figure \ref{fig:rate_1}. The etch depth is measured after the anisotropic etch step to ensure that the process has not drifted. 

\begin{figure}
    \centering
    \includegraphics[width=1\linewidth]{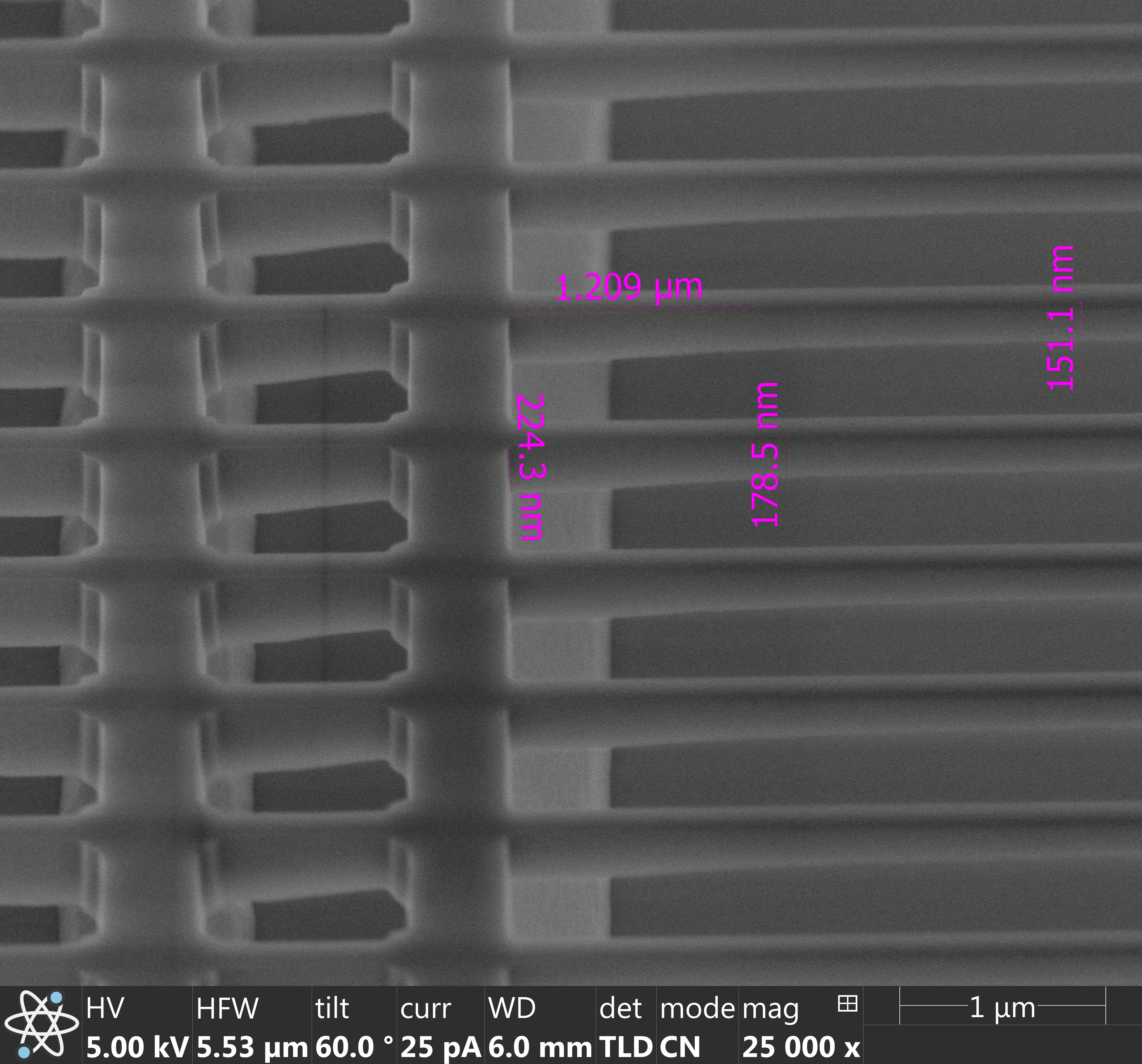}
    \caption{SEM image of a membrane showing the variation of thickness uniformity across the nanobeams. They become effectively uniform at distances > ~1 µm from the trellis edge (see the labels of the corresponding dimensions}
    \label{fig:Uniformity_1}
\end{figure}

The symmetry inherent in the nanobeam membrane leads to an arch-like region in the first 1.2µm of any given nanobeam (Figure \ref{fig:Uniformity_1}). After this region, the nanobeams are uniform within the tolerances of our SEM measurements. 

\begin{figure}
    \centering
    \includegraphics[width=1\linewidth]{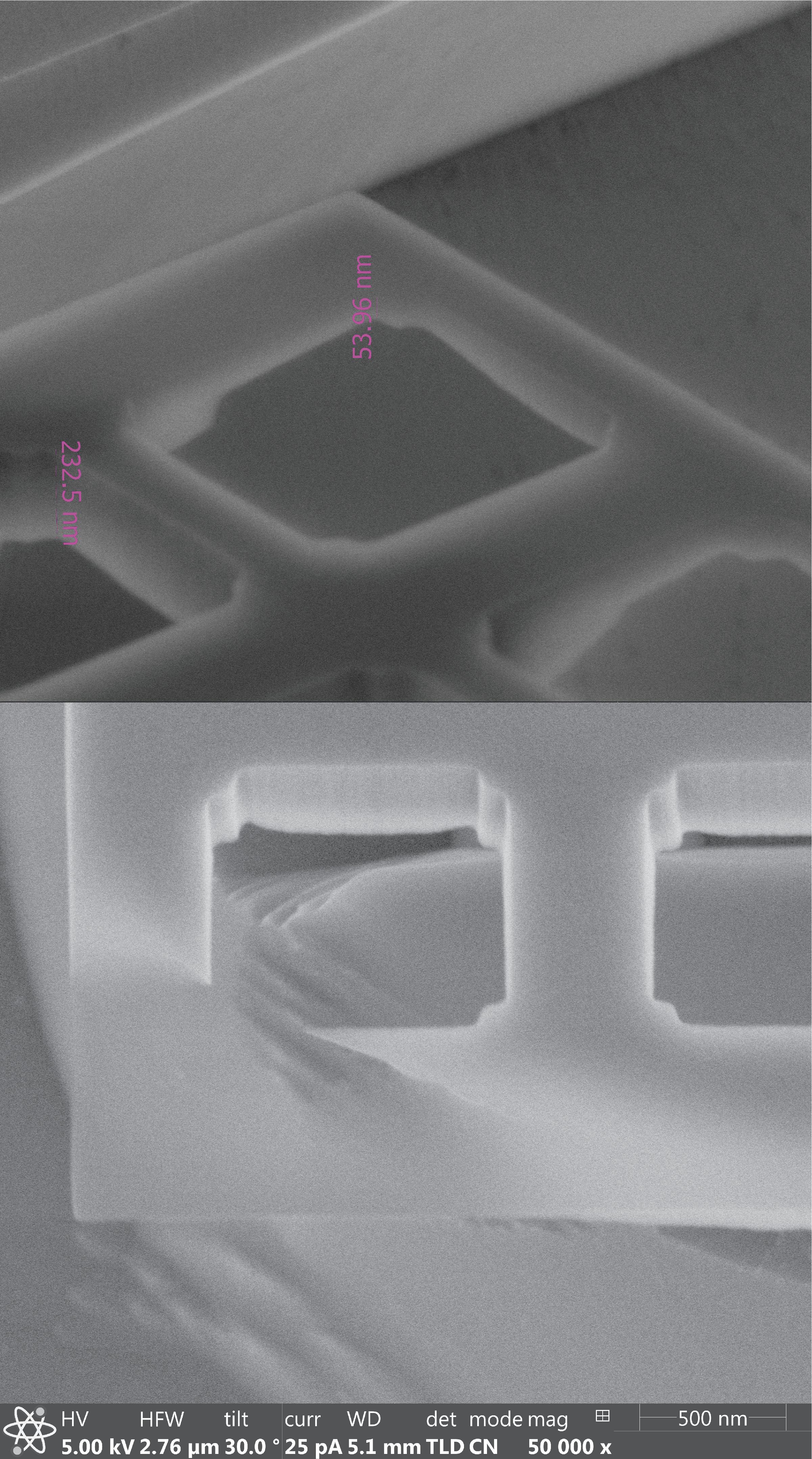}
    \caption{SEM images of the corner of a frame showing extreme thinning. The trellis has a redundant layer to buffer for this effect and to keep the nanobeams tied together as a membrane. Top and bottom images are of different corners on different frames.}
    \label{fig:thinning_1}
\end{figure}

The outer parts of the frames experience different area dependent etching. On the corners of the frames there is extreme thinning. To maintain frame integrity, for thin membranes, a second layer was added to the frame to build the 'trellis' (Figure \ref{fig:thinning_1}).  

\section{Experimental setup for confocal microscopy}
All the spin and charge state measurements (Figure \ref{fig:Figure2} and \ref{fig:Figure3}) are carried out on a home-built confocal microscope that uses an oil-immersion objective (Nikon 100x, NA = 1.3) for excitation and photoluminescence collection. The sample is mounted on a glass slide with an omega-shaped antenna deposited on it that is used for microwave delivery. Green excitation is provided by a 520 nm modulated laser diode (Cobolt MLD-06), orange excitation is provided by a 591 nm HeNe laser (Newport) modulated via acoustic-optic modulator (Isomet). The lasers are combined into a single path via an RGB fiber combiner (Thorlabs RGB46HF) with a 650 nm shortpass optical filter (Thorlabs FELH650) in the path. A 4f lens system with scanning galvo mirrors (Thorlabs) is used for scanning the beam across the field of view. A 650 nm longpass dichroic mirror is used to separate collection from excitation. The collection is coupled into a silicon avalanche photodiode (Excelitas) via a single mode fiber (Thorlabs 630HP) and a 650 nm longpass optical filter (Thorlabs) in the path. The microwave (MW) pulses are generated by a SRS SG384 signal generator with an I/Q modulation controlled by an Arbitrary Waveform Generator (Keysight 33600A). A MW switch (Mini Circuits ZASWA-2-50-DRA+) is used to modulate the MW signal, which is then amplified (Mini Circuits ZHL-16W-43-S+). All the TTL gates are provided by a Pulseblaster (SpinCore).

Photoluminescence saturation experiments (Figure \ref{fig:Figure4}) are performed on a home-built confocal air-objective microscope equipped with a 3D scanning stage (MadCity Labs). A 520 nm modulated laser diode (Cobolt MLD-06) is used for excitation. A 650 nm longpass dichroic mirror is used to separate collection from excitation. The collection is coupled into a silicon avalanche photodiode (Excelitas) via a single mode fiber (Thorlabs 630HP) and a 650 nm longpass optical filter (Thorlabs FELH650) in the path. 

\section{Experimental details of NV center spin and charge measurements}

\begin{figure}
    \centering
    \includegraphics[width=1\linewidth]{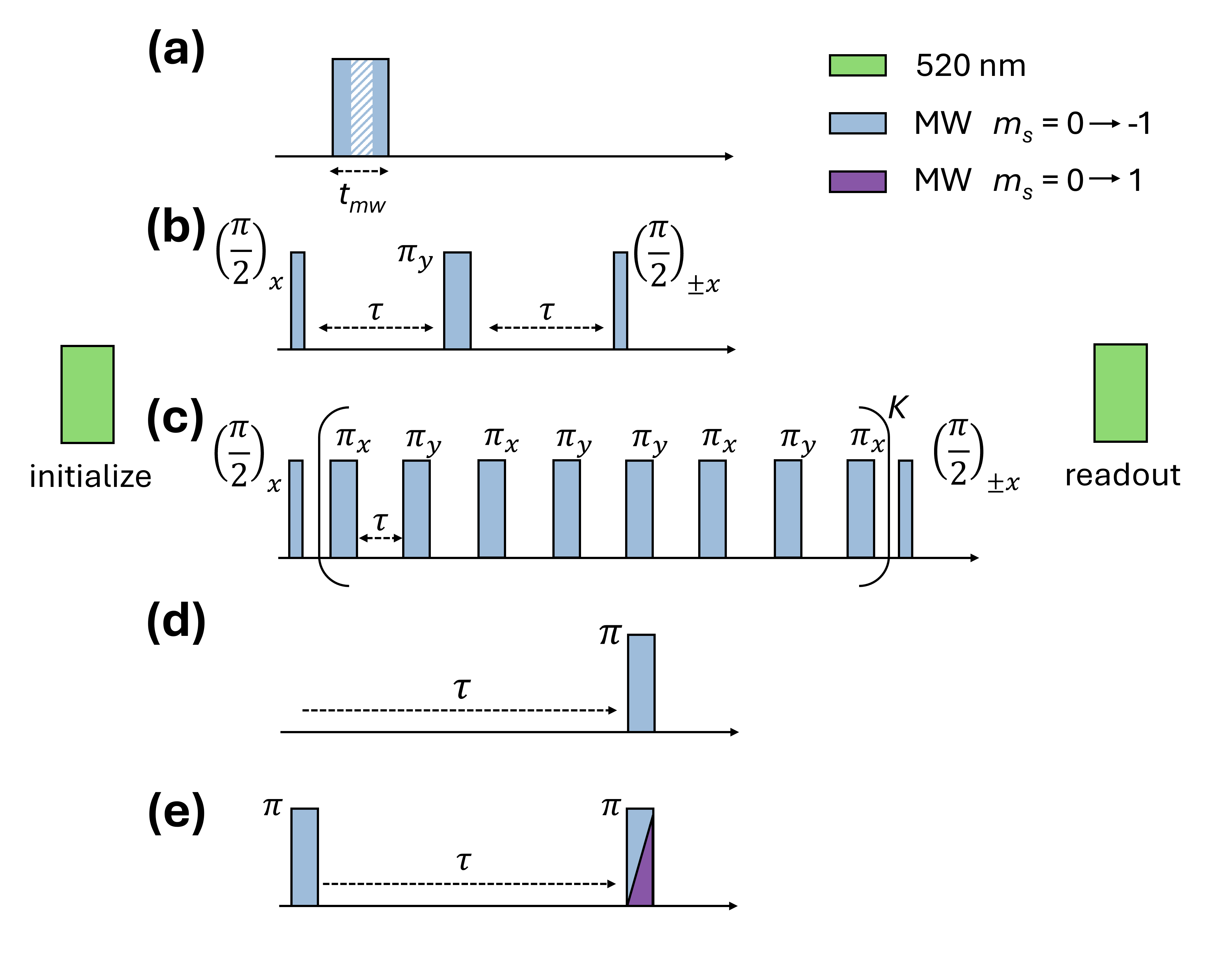}
    \caption{Pulse sequences used for the experiments presented in Section 2.2 and 2.3: (a) Rabi oscillations, (b) Hahn echo, (c) XY8-K dynamical decoupling, (d) Single-quantum $T_1$ relaxometry. (e) Double-quantum $T_1$ relaxometry with the last MW pulse alternating between the two spin transitions ($m_s = 0  -> -1$ and $m_s = 0  -> 1$). Green 520 nm illumination is used for both initialization and readout.}
    \label{fig:Supp1s}
\end{figure}

Spin experiment results shown in the main text use the pulse sequences detailed in Figure \ref{fig:Supp1s}. In every experiment, we use off-resonant green (520 nm) excitation to initialize and readout the spin state of the NV center ground triplet. Figure \ref{fig:Supp1s}a is a Rabi experiment used to determine the spin contrast shown in Figure \ref{fig:Figure3}d. Figure \ref{fig:Supp1s}b shows a Hahn echo experiment used to determine $T_2$ decoherence times presented in Figure \ref{fig:Figure2}a. Figure \ref{fig:Supp1s}c describes an XY8-K dynamical decoupling sequence used to obtain the data in Figure \ref{fig:Figure2}b,c. Finally, Figure \ref{fig:Supp1s}d shows a pulse sequence for a single-quantum and double-quantum $T_1$ relaxometry. Here, an additional microwave pulse exciting $m_s$=0 to 1 ground triplet transition is used to probe the double-quantum $T_1$. A separate signal generator (SRS SG384) and a microwave switch (Mini Circuits ZASWA-2-50-DRA+) is combined with the original microwave line using an RF-combiner (Mini-Circuits ZFRSC-42-S+). 
All the data in Figure \ref{fig:Figure2}a-c was taken in the aligned magnetic field of 360 Gauss provided by a permanent magnet. Single- and double-quantum $T_1$ data was taken at 35 Gauss. Spin contrast data in Figure \textbf{\ref{fig:Figure3}}d was taken at 25 Gauss. 

\begin{figure}
    \centering
    \includegraphics[width=1\linewidth]{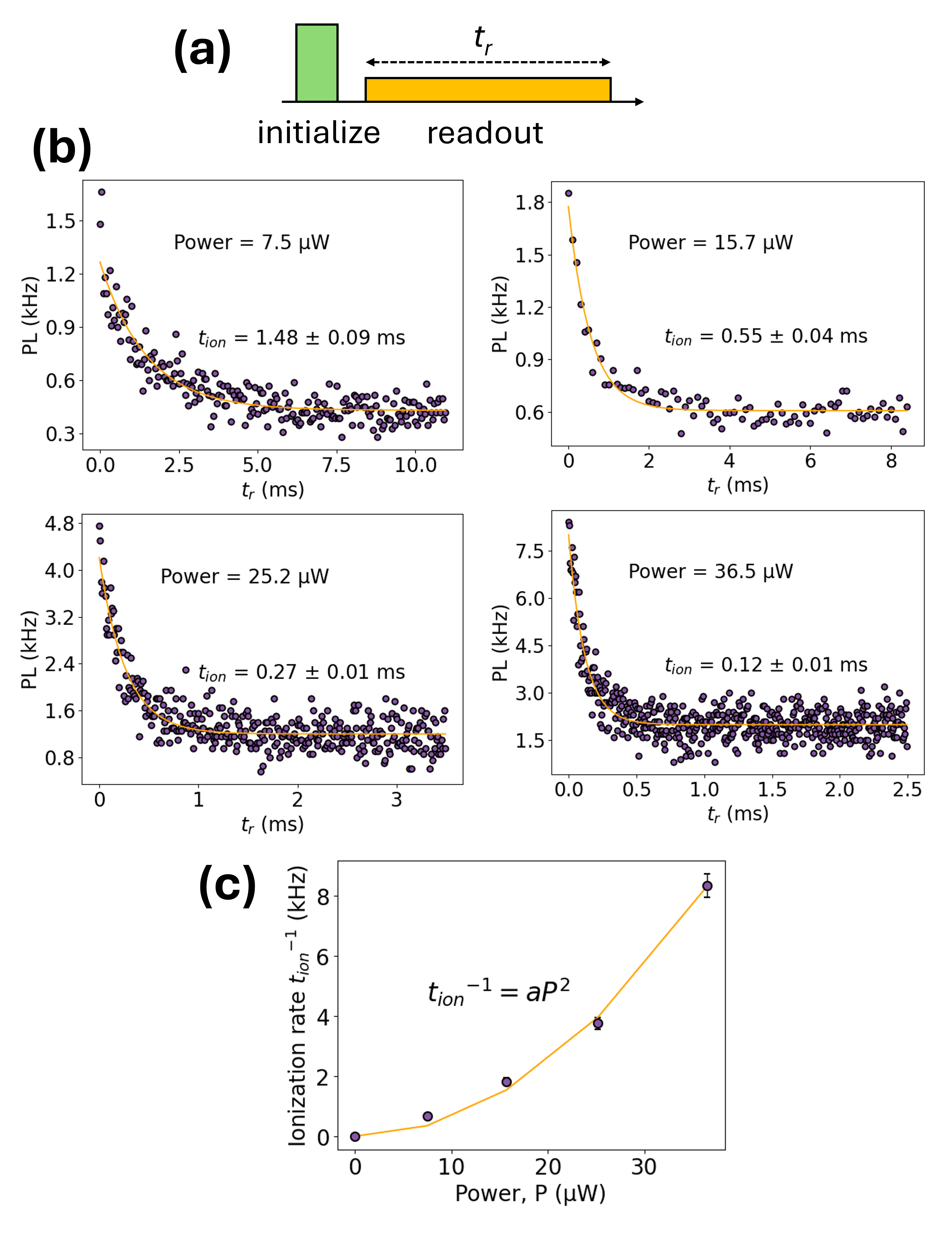}
    \caption{(a) Pulse sequence used for charge state and ionization rate measurements. 520 nm green illumination initializes NV$^-$, a weak 594 nm orange illumination interrogates the charge state population. (b) NV$^-$ photoluminescence decay under orange illumination of variable power fitted with an exponential function. (c) Fitted rates plotted as a function of orange power. Orange line is a quadratic fit to the data: ${t_{ion}}^{-1} = aP^2$}
    \label{fig:Supp2s}
\end{figure}

To reconstruct the spectral noise density measured by an XY8 pulse sequence, we use the approach described previously in [31]. Decoherence decay signal $C(T)$ for $T = N\tau$ where $\tau$ is $\pi$ pulse separation and $N$ is a number of $\pi$ pulses in the XY8 sequence ($N = 8*K$ in Figure \ref{fig:Supp2s}c) can be written as:
\begin{equation}
C(T) = e^{-\chi(t)},
\label{eqS1}
\end{equation}
where $\chi(t)$ is a phase acquired due to fluctuating magnetic fields. It is connected to the noise spectral density $S(\omega)$ through a filter function determined by the pulse sequence. It is typical to approximate this filter function as a $\delta$-function that peaks at $\omega=\frac{\pi N}{T}$, which leads to a straightforward relation between the measured signal $C(T)$ and $S(\omega)$:
\begin{equation}
S(\omega) = -\frac{\pi log(C(T))}{T},
\label{eqS2}
\end{equation}
Charge state population measurements shown in Figure \ref{fig:Figure3}b are carried out using a pulse sequence in Figure \ref{fig:Supp2s}(a) with a 594 nm orange readout at 7 $\mu$W at a fixed $t_r$ for each NV center. Ionization under orange is measured by varying $t_r$ at a fixed optical power and recording photoluminescence decays shown in Figure \ref{fig:Supp2s}(b), which are fit to an exponential function. The corresponding decay rates are then plotted as a function of power and are fit to a quadratic polynomial $aP^2$ (Figure \ref{fig:Supp2s}(c)). Ionization rate coefficients $a$ are then plotted as a distribution in Figure \ref{fig:Figure3}(c).

\begin{figure}
    \centering
    \includegraphics[width=1\linewidth]{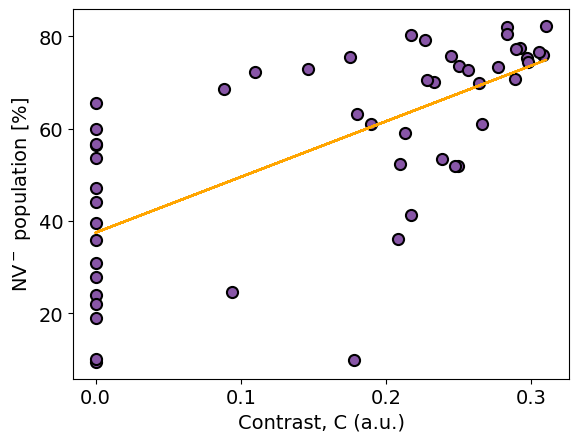}
    \caption{NV$^-$ population as a function of spin contrast measured using Rabi oscillation experiment for 53 NV centers in the nanobeams (the same subset as in Figure 3(d)). Orange line is a linear fit showing a correlation between the spin contrast and the charge state population. Most of the NV centers with no spin contrast also have low NV$^-$ population.}
    \label{fig:Supp3s}
\end{figure}

In Figure \ref{fig:Supp3s}, we plot NV$^-$ population as a function of spin contrast measured using Rabi oscillations for the same 53 NVs in the nanobeams, which are interrogated in Figure \ref{fig:Figure3}(d). Linear fit shows correlation between lower contrast and bad charge state, which explains slightly increased prevalence of NV centers with no spin contrast in this subset compared to the unprocessed host diamond.

\section{FDTD Simulations of NV Centers in Nanobeams}\label{supp-fdtd}

Simulations performed in Lumerical are performed on a single diamond ($n_{diamond}$=2.4) nanobeam 30 µm in length with an NV center nominally placed at the nanobeam center. Collection under a 0.9 NA is simulates the emission of an electric dipole at the NV center position. The collection efficiency is calculated by integrating over the wavelength range of the NV center emission that reaches the objective. The collection efficiency of a given NV center in a nanobeam depends strongly on the NV center depth (Figure \ref{fig:Supp2}), dipole orientation (Figure \ref{fig:Supp3}), lateral position (Figure \ref{fig:Supp4}), and transfer orientation (flipped or unflipped, Figure \ref{fig:Figure4}b). We note that the collection efficiency is not optimized for a centered NV center; the dipole orientation of the NV center shifts the emission profile, causing some offset to be desirable to achieve maximum vertical transmission. 

\begin{figure}
    \centering
    \includegraphics[width=1\linewidth]{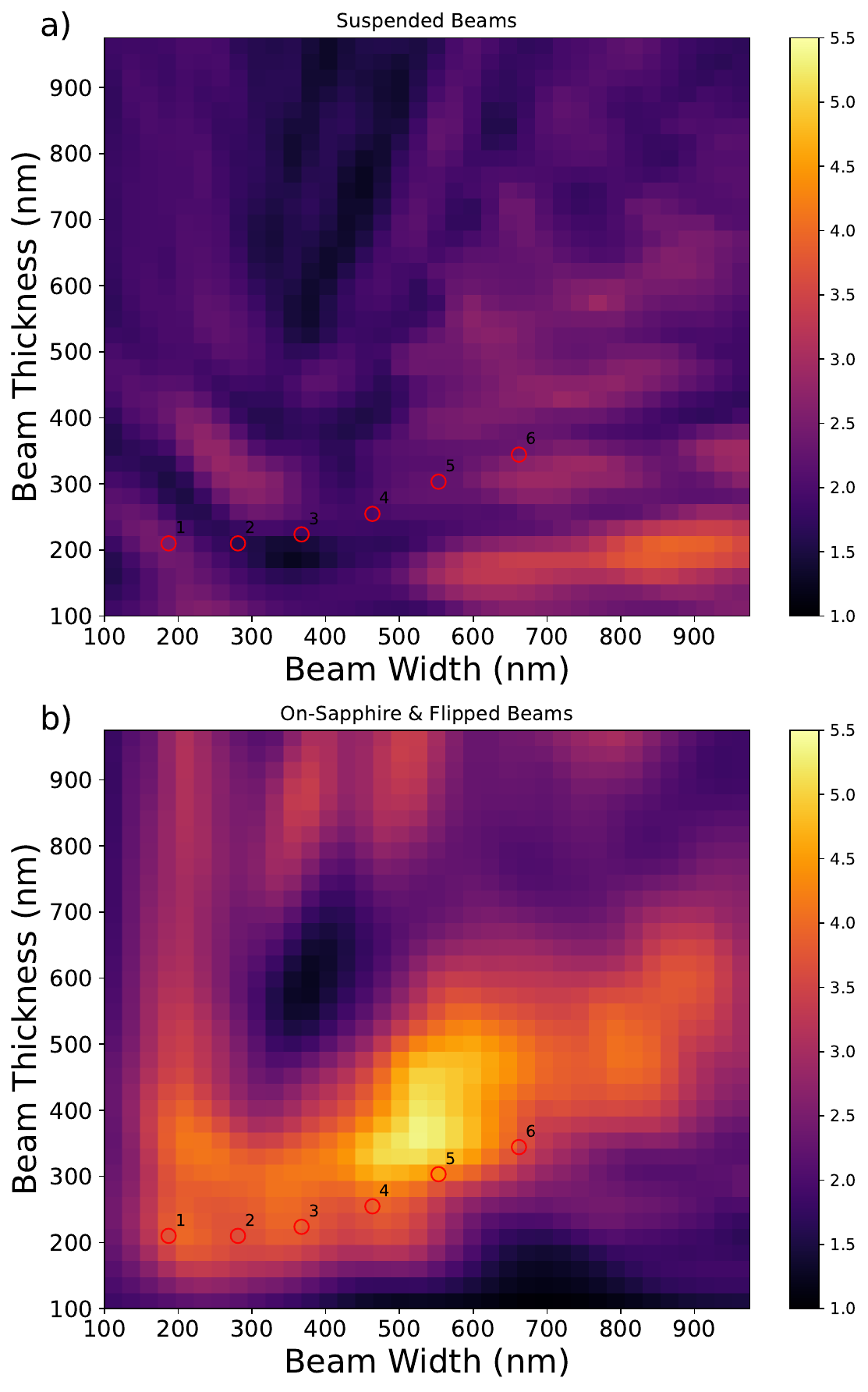}
    \caption{(a) FDTD simulation of collection efficiency (CE) enhancement for a range of nanobeam geometries, both when suspended and (b) when flipped onto sapphire. Each point represents the average over all four NV center orientations, for an NV center centered laterally within the nanobeam. Red circles represent the six frame geometries we fabricated in the main text. The thickness variation in width highlights the aspect ratio dependent etch rate of the quasi-isotropic etch. We specifically targeted our fabrication to create nanobeams around the maximum for CE around 550 nm wide and 350 nm thick. }
    \label{fig:Supp1}
\end{figure}

\begin{figure}
    \centering
    \includegraphics[width=1\linewidth]{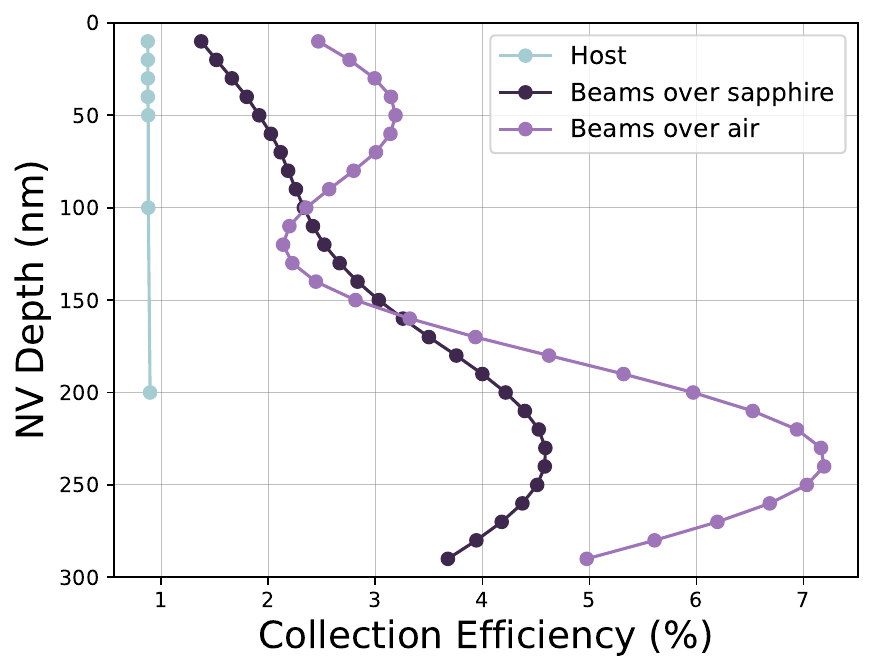}
    \caption{The collection efficiency as a function of NV center z-position, showcasing a negligible dependence in the unfabricated host crystal as compared to the strong dependence in a nanobeam. In particular, CE generally increases with depth in the nanobeam, indicating increased interactions with the nanobeam sidewalls, which serve to guide light vertically toward the objective. As a result, NV centers positioned at the bottom of the nanobeam (the situation after we flip a membrane), are much brighter than those at the top.}
    \label{fig:Supp2}
\end{figure}

\begin{figure}
    \centering
    \includegraphics[width=1\linewidth]{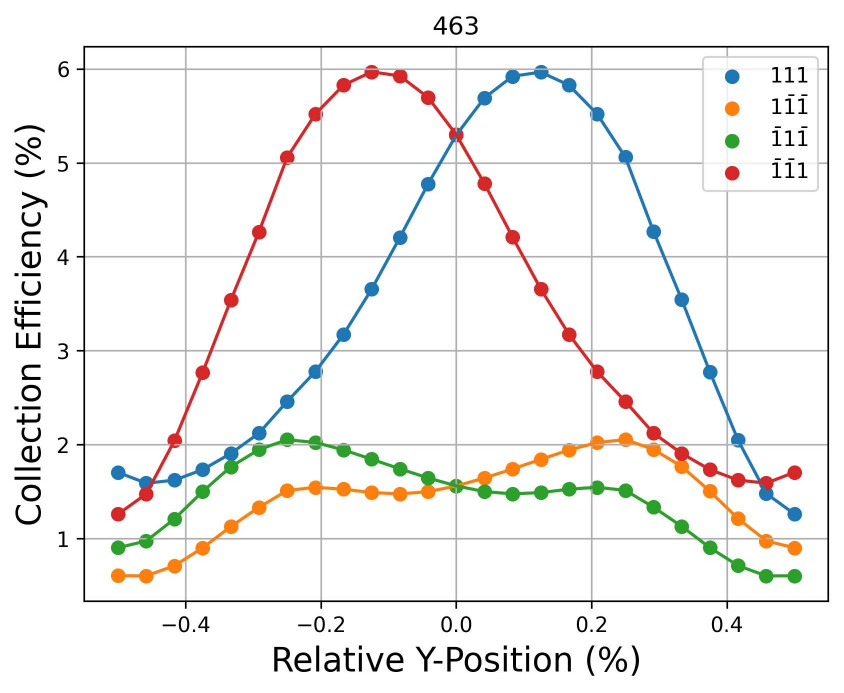}
    \caption{The collection efficiency strongly depends both on dipole orientation and lateral position of the NV center in y, relative to the edge of the nanobeam. This simulation is for a nanobeam that is 463 nm wide and 250 nm thick. As NV center dipole orientation is not symmetric with respect to our nanobeam geometry, different geometries favor different lateral positions, typically off-center. Two of the dipole orientations are oriented such that they are not angled toward our objective, and collection from these orientations is much weaker than those facing toward the objective. }
    \label{fig:Supp3}
\end{figure}

\begin{figure}
    \centering
    \includegraphics[width=1\linewidth]{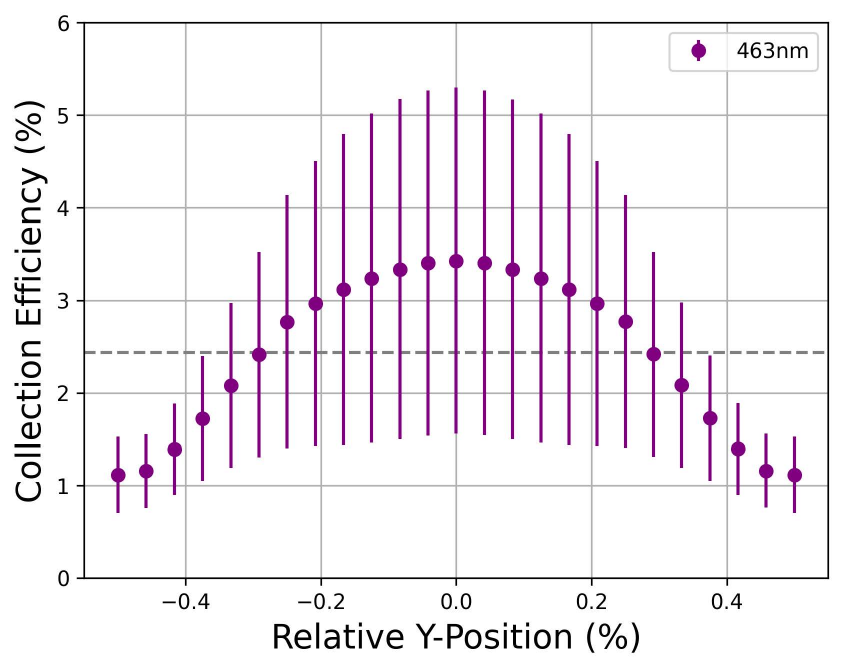}
    \caption{Simulated collection efficiency as a function of relative lateral displacement, averaging over all four orientations. Error bars represent the standard deviation resulting from the four orientations.}
    \label{fig:Supp4}
\end{figure}

For a laterally centered NV center, we also sweep nanobeam geometry to maximize collection efficiency. We define the x-axis as the direction along the nanobeam, and the y-axis as toward the nanobeam edge. For suspended nanobeams, we observe no singular "best" case nanobeam geometry (Figure \ref{fig:Supp1}a), though some preference exists for wide, but thin nanobeams. In contrast, when nanobeams are flipped and on sapphire Figure (\ref{fig:Supp1}b), a clear region exists where the average enhancement over all dipole orientations is maximized. We fabricate nanobeams targeting this region, which is highlighted by the red circles in Figure \ref{fig:Supp1}. Due to the vertical etch rate of the quasi-isotropic etch, less wide nanobeams are also thinner (in thickness), causing the observed variation in our fabricated nanobeam geometries.

\begin{figure}
    \centering
    \includegraphics[width=1\linewidth]{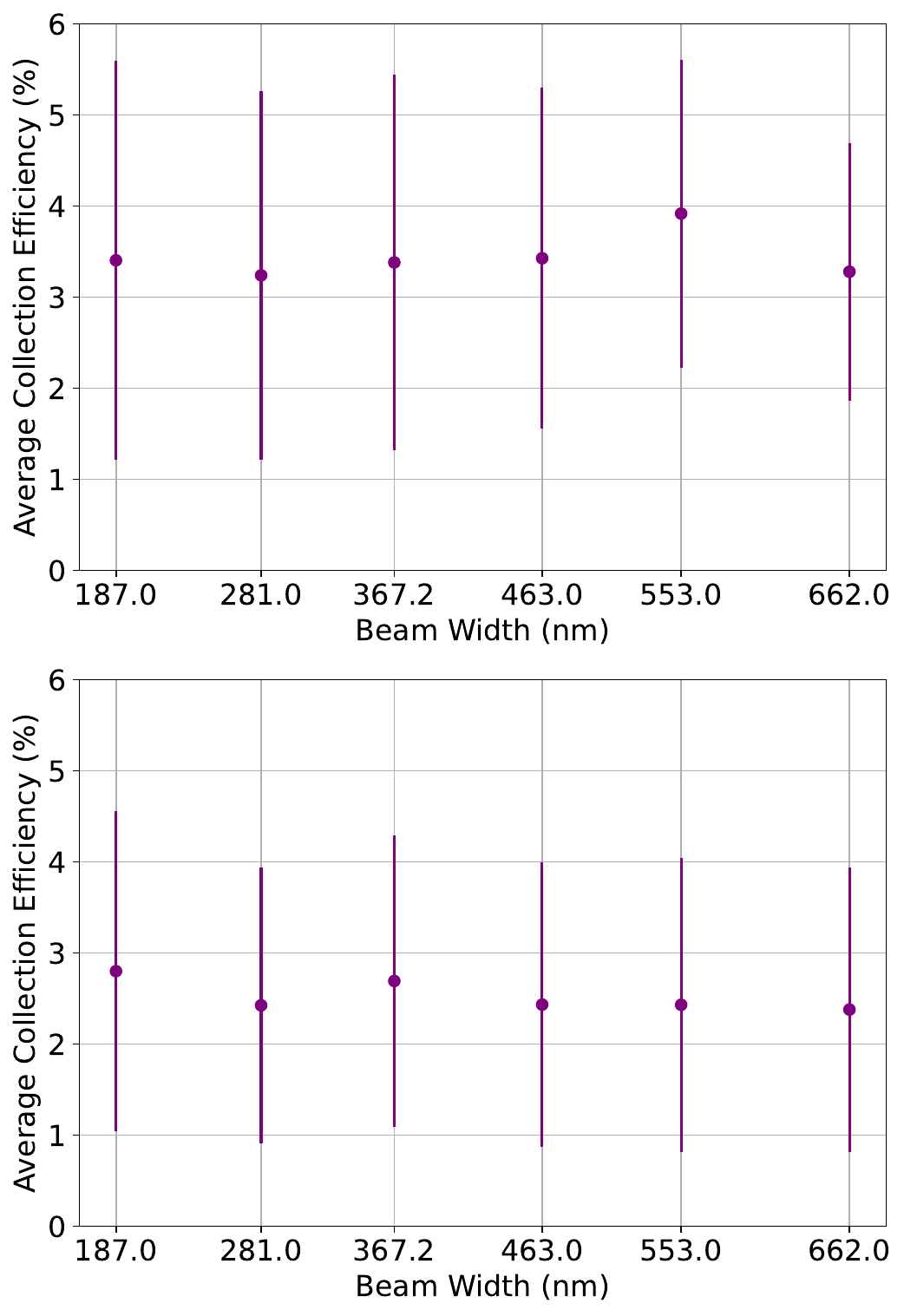}
    \caption{(a) Average CE for a given nanobeam geometry over all dipole orientations and lateral positions. (b) Average collection for a given nanobeam geometry over all dipole orientations, but only for NV centers centered laterally. Accounting for lateral displacement lowers the average collection efficiency in a beam, but the spread is still large (dominated by the dipole orientation dependence).}
    \label{fig:Supp5}
\end{figure}

\end{document}